\documentclass[11pt,a4paper]{article}
\pdfoutput=1
\usepackage{jheppub}
\usepackage{rotating}
\usepackage{array}
\usepackage{amsmath}
\usepackage{subcaption}
\usepackage[normalem]{ulem}
\usepackage{slashed}
\usepackage{booktabs}
\usepackage[pdftex,table,dvipsnames]{xcolor}
\usepackage[utf8]{inputenc}
\usepackage{units}
\usepackage{multirow}
\usepackage{verbatim}
\usepackage{url}
\usepackage{xspace}
\usepackage{color}
\preprint{TTK-20-08}

\title{On the interplay between astrophysical and laboratory probes of MeV-scale axion-like particles}

\author{Fatih Ertas and Felix Kahlhoefer}

\affiliation{ Institute for Theoretical Particle Physics and Cosmology (TTK), RWTH Aachen University, D-52056 Aachen, Germany}

\emailAdd{ertas@physik.rwth-aachen.de}
\emailAdd{kahlhoefer@physik.rwth-aachen.de}

\abstract{Studies of axion-like particles (ALPs) commonly focus on a single type of interaction, for example couplings only to photons. Most ALP models however predict correlations between different couplings, which change the phenomenology in important ways. For example, an MeV-scale ALP coupled to Standard Model gauge bosons at high energies will in general interact with photons, $W^\pm$ and $Z$ bosons as well as mesons and nucleons at low energies. We study the implications of such scenarios and point out that astrophysical constraints, in particular from SN1987A, may be substantially relaxed, opening up new regions of parameter space that may be explored with laboratory experiments such as NA62.}


\begin{document}

\maketitle

\flushbottom

\section{Introduction}
Axions, which were first introduced as a solution to the strong CP problem~\cite{Peccei:1977hh, Peccei:1977ur,Weinberg:1977ma,Wilczek:1977pj}, have since been studied more generally as potential new pseudoscalar particles with small mass and tiny interaction rates with the Standard Model (SM) particle content. These candidates are then commonly referred to as Axion-like particles (ALPs). What makes them in particular intriguing is that they can arise as Pseudo-Goldstone bosons of a spontaneously broken approximate global symmetry explaining both their small mass and couplings. Since the ALP coupling strength to SM particles as well as their precise cosmological history are model-dependent, they have been studied in literature in a wide phenomenological range, e.g.~in the context of dark matter~\cite{Dolan:2014ska,Boehm:2014hva,Tunney:2017yfp,Arcadi:2017wqi,Kahlhoefer:2017umn,Abe:2018emu,Ertas:2019dew,Daido:2019tbm,Alonso-Alvarez:2019ssa,Arcadi:2020gge,Hochberg:2018rjs}, flavour physics~\cite{Choi:2017gpf,Dobrich:2018jyi,Cornella:2019uxs,Bauer:2019gfk,MartinCamalich:2020dfe}, astrophysics~\cite{Vysotsky:1978dc,Raffelt:1996wa,Raffelt:2006cw,Lee:2018lcj,Chang:2018rso,Jaeckel:2019xpa} and cosmology~\cite{Cadamuro:2011fd,Millea:2015qra,Graham:2015cka,Hoof:2017ibo,Hoof:2018ieb,Depta:2020wmr}. Consequently, there are many experimental efforts to search for different ALP candidates, see e.g.~refs.~\cite{Arik:2008mq,Bahre:2013ywa,Armengaud:2014gea,Alekhin:2015byh,Graham:2015ouw,Dobrich:2015jyk,Berlin:2018pwi,Curtin:2018mvb,Berlin:2018bsc,Ariga:2018uku}.

As ALPs naturally emerge from the breaking of a global symmetry at some high scale they can be suitably studied within an effective field theory (EFT) approach~\cite{Georgi:1986df,Bauer:2017ris,Brivio:2017ije}. In such a set-up many different interactions are expected to arise in a correlated way leading to a rich phenomenology. However, most existing ALP studies limit themselves to a single kind of interaction, most prominently the coupling to photons, while all other couplings are assumed to be zero or at least sufficiently small. In this case it is not possible to capture all of the phenomenological aspects of an ALP that has multiple interactions at low energies.

The present work therefore combines and improves various approaches for constraining ALPs from the literature:
\begin{itemize}
 \item For a given ALP interaction at high scales we consider the effect of all induced low-energy interactions. For example, if the ALP couples to SM $SU(2)_L$ gauge bosons we do not only include the resulting $W$ boson interaction but also the effective ALP-photon coupling and the $Z \gamma a$ vertex. In the case of ALP-gluon interactions we include the induced ALP-photon coupling together with the nucleon interactions.
 \item This approach makes it essential to go beyond the single-operator-study usually employed in the ALP literature (see also refs.~\cite{Gavela:2019wzg,Alonso-Alvarez:2018irt} for similar recent approaches). We consistently include all relevant effects by taking into account all required operators in both the ALP production and its decay or absorption. This treatment also differs from refs.~\cite{Izaguirre:2016dfi,Gavela:2019wzg}, where it is assumed that the ALPs decay invisibly into some additional light particles.
 \item We consistently include all relevant interactions to determine the bounds on ALP-photon and ALP-gluon couplings from the cooling of supernova SN1987A. A similar analysis has been performed in ref.~\cite{Lee:2018lcj}, but we substantially improve the calculation of the ALP optical depth, of the bounds on ALP-nucleon interactions (following ref.~\cite{Chang:2018rso} in both cases) and of the estimation of uncertainties affecting the constraints. Compared to previous results, our main constraint on the ALP-photon coupling is more conservative in the trapping regime whereas our bound on ALP-gluon interactions excludes more parameter space for large ALP masses.
\end{itemize}

We apply this approach to the case of ALPs with MeV-scale masses that interact with the various SM gauge bosons. We are particularly interested in the question whether such ALPs may be discovered with laboratory experiments. The most promising search channel is the rare decay $K^+ \to \pi^+ + \text{inv}$, which will be explored with unprecedented sensitivity at NA62~\cite{NA62:2017rwk}. In the simplest models, for example in the case of an ALP coupling only to $SU(2)_L$ gauge bosons, the parameter region probed by NA62 is already excluded by astrophysical constraints, in particular by the bound from SN1987A. However, as soon as interactions with several different gauge bosons are considered simultaneously, supernova bounds can be strongly suppressed. The reason is that for sufficiently large couplings ALPs are trapped in the supernova core and can no longer contribute significantly to its cooling. We investigate this interplay of astrophysical and laboratory probes in detail and find that NA62 has great potential to explore this so-called trapping regime.

The paper is organized as follows. In section~\ref{sec:TheoFrame} we review the general aspects of the ALP EFT and derive the relevant interactions as well as meson-mixing contributions. We then discuss existing constraints and prospects on MeV-scale ALPs paying particular attention to supernova bounds in section~\ref{sec:ExpConstr}. Our results for specific coupling scenarios of ALPs to SM gauge boson are presented in section~\ref{sec:results}, which is then followed by our conclusions in section~\ref{sec:conclusions}. More details on the mixing between ALPs and pseudoscalar mesons as well as their contributions to $K^+ \rightarrow \pi^+ + \text{inv.}$ are given in appendices~\ref{app:ChiralInt} and~\ref{app:KtoPiMix}. Finally, we give details on the computation of the ALP luminosity for SN1987A in appendix~\ref{app:LumDetails}, review the influence of uncertainties on the SN bounds in appendix~\ref{app:SNUncer} and discuss the sensitivity of NA62 to ALPs with mass greater than the pion mass in appendix~\ref{app:NA62SR2}.

\section{Theoretical framework}
\label{sec:TheoFrame}
\subsection{ALP effective theory}
\label{subsec:ALPEFT}
In this work we employ an EFT approach to describe the interactions of ALPs with SM particles at energies below the new-physics scale $\Lambda$. For this we assume that the ALP $a$ is a singlet under the SM gauge group, CP odd and that its mass is below the electroweak scale. We additionally assume no new sources of CP violation. Following the notation from ref.~\cite{Bauer:2017ris}, the Lagrangian then reads~\cite{Georgi:1986df,Brivio:2017ije}
\begin{align}
 \begin{split}
\label{eq:ALPeffLagr}
 \mathcal{L}_\text{eff} =& \frac{1}{2} (\partial_\mu a)(\partial^\mu a) - \frac{m_{a,0}^2}{2} a^2 + \frac{\partial^\mu a}{\Lambda} \sum_F \bar{\Psi}_F \, C_F\, \gamma_\mu\, \Psi_F\\
&+g_s^2\, C_{GG} \frac{a}{\Lambda} G^a_{\mu\nu} \tilde{G}^{\mu\nu,a}+g^2\,C_{WW} \frac{a}{\Lambda} W^a_{\mu\nu} \tilde{W}^{\mu\nu,a} + g'^{\,2} \,C_{BB} \frac{a}{\Lambda} B_{\mu\nu} \tilde{B}^{\mu\nu}\,,
\end{split}
\end{align}
where $G^a_{\mu\nu}$, $ W^a_{\mu\nu}$ and $B_{\mu\nu}$ describe the field strength tensors of the strong and electroweak sector with $C_{GG}$, $C_{WW}$ and $C_{BB}$ being the corresponding Wilson coefficients and $g_s$, $g$ and $g'$ are the SM gauge couplings. The dual tensors are given by $\tilde{X}^{\mu\nu} = \frac{1}{2} \epsilon^{\mu\nu\alpha\beta} X_{\alpha\beta}$ (\mbox{$\epsilon^{0123} = 1$}). Furthermore, $\Psi_F$ denote the SM fermion multiplets and $C_F$ are (hermitian) matrices characterizing their couplings strength to the ALP field. In the following we will focus on ALPs interacting predominantly with SM gauge bosons and therefore assume $C_F \approx 0$ at tree level.

The continuous shift symmetry of the ALP, $a \rightarrow a + \text{const.}$, used to construct the Lagrangian above is broken down to a discrete one by the anomalous $G^a_{\mu\nu} \tilde{G}^{\mu\nu,a}$ interactions and is furthermore softly broken by the explicit ALP mass term $m_{a,0}$. Combining the influence of these two terms on the ALP mass, we obtain~\cite{Bauer:2017ris,Weinberg:1977ma}
\begin{align}
 \begin{split}
 m^2_a &\approx m_{a,0}^2 +(32\,\pi^2 |C_{GG}|)^2 \,\frac{F_0^2 m_\pi^2}{\Lambda^2}\,
 \frac{m_u m_d}{(m_u + m_d)^2}\\
 &\approx m_{a,0}^2 + \left(1.8\,\text{MeV}\,|C_{GG}|\,\frac{1\,\text{TeV}}{\Lambda}\right)^2\,,
 \end{split}
 \end{align}
where $F_0 \approx 92\,\mathrm{MeV}$ is the pion decay constant. The parameter $m_{a,0}$ therefore allows us to treat the ALP mass and the ALP-gluon coupling as independent parameters of our theory, in contrast to the case of the QCD axion~\cite{Peccei:1977hh, Peccei:1977ur,Weinberg:1977ma,Wilczek:1977pj}.

The next step is to include the effects of electroweak symmetry breaking (EWSB) into the effective theory for ALPs. This induces several new interactions in the electroweak sector, of which the following ones are most relevant to our discussion~\cite{Bauer:2017ris}
\begin{align}
 \begin{split}
 \mathcal{L}_\text{eff} \supset\, e^2\,C_{\gamma\gamma} \frac{a}{\Lambda} F_{\mu\nu} \tilde{F}_{\mu\nu} + \frac{2 e^2}{s_w c_w} C_{\gamma Z} \frac{a}{\Lambda} F_{\mu\nu} \tilde{Z}_{\mu\nu} + 4 g^{\,2} \,C_{WW} \frac{a}{\Lambda} \epsilon^{\mu\nu\alpha\beta} \partial_\mu W^+_\nu \partial_\alpha W^-_\beta \,,
\end{split}
\end{align}
where 
\begin{align}
\label{eq:EWCoupl}
 C_{\gamma\gamma} = C_{WW} + C_{BB}\,,\qquad C_{\gamma Z} = c^2_w C_{WW} - s_w^2 C_{BB}\,.
\end{align}
Here $e$ is the electric charge, $\theta_w$ is the weak angle with $s_w \equiv \sin(\theta_w)$ and $c_w \equiv \cos(\theta_w)$, $F^{\mu\nu}$ and $Z^{\mu\nu}$ are the photon and Z boson field strength tensors, respectively, and $W^{\pm}$ are the fields for the $W^\pm$ bosons.

\subsection{Mixing effects of light ALPs}
\label{subsec:Mixing}
As we are interested in studying ALPs with masses below 1\,GeV, the hadronic contributions to the processes of interest (discussed in more detail later) can be obtained from a chiral Lagrangian~\cite{Georgi:1986df}. We start by computing the ALP-mixing with pseudoscalar mesons ($\pi^0$, $\eta$, $\eta'$). For this we first have to determine the interactions resulting from eq.~\eqref{eq:ALPeffLagr} at scales of about 1\,GeV~\cite{Bauer:2017ris}
\begin{align}
 \begin{split}
 \mathcal{L}_\text{eff}\supset& \frac{1}{2} (\partial_\mu a)(\partial^\mu a) - \frac{m_{a,0}^2}{2} a^2 +g_s^2\, C_{GG} \frac{a}{\Lambda} G^a_{\mu\nu} \tilde{G}^{\mu\nu,a}\\
 &+  e^2\,C_{\gamma\gamma} \frac{a}{\Lambda} F_{\mu\nu} \tilde{F}^{\mu\nu}+\bar{q}\,(i\slashed{D} - M_q)q\,,
 \end{split}
\end{align}
where we have included the relevant QCD terms with $q = (u,\,d,\,s)^{T}$ and the quark mass matrix $M_q = \text{diag}(m_u,\,m_d,\,m_s)$. 

Before mapping this onto a chiral Lagrangian, it is convenient to rotate the gluonic interaction away by performing a chiral rotation on the light quark fields $q \rightarrow e^{i \kappa_q \frac{a}{2f_a} \gamma_5} q$ where $\text{Tr}[\kappa_q]=1$ and $1/f_a = -32 \pi^2 C_{GG}/\Lambda$. The interactions then read~\cite{Bauer:2017ris,Georgi:1986df}
\begin{align}
 \begin{split}
\label{eq:LagrChirRot}
  \mathcal{L}_\text{eff} \supset &\frac{1}{2} (\partial_\mu a)(\partial^\mu a) - \frac{m_{a,0}^2}{2} a^2  +e^2 \left(C_{\gamma\gamma} - 6 C_{GG} \text{Tr}[\kappa_q Q^2_q] \right) \frac{a}{\Lambda} F_{\mu\nu} \tilde{F}_{\mu\nu}\\
  &+\bar{q}\,(i\slashed{D} - M_q(a))q+ \frac{\partial^\mu a}{2\Lambda} \bar{q}\left( (\kappa_q 32 \pi^2 C_{GG}) \gamma_\mu \gamma_5 \right)q\,.
  \end{split}
\end{align}
Here we introduced the rotated quark mass matrix $M_q(a) =e^{i \kappa_q \frac{a}{2f_a} \gamma_5} M_q e^{i \kappa_q \frac{a}{2f_a}\gamma_5}$ and the charge matrix $Q_q = \text{diag}(2/3,\,-1/3,\,-1/3)$ containing the electric charges of the light quarks in units of $e>0$.

These interactions can now be mapped onto a chiral Lagrangian at leading order (see appendix \ref{app:ChiralInt} for details). Here it is convenient to choose $\kappa_q = M_q^{-1}/\text{Tr}[M_q^{-1}] $ in order to avoid tree-level mass mixing between the ALP and the octet mesons~\cite{Georgi:1986df}. The mixing contributions of the ALP into the pseudoscalar mesons can then be determined to~\cite{Aloni:2018vki}
\begin{align}
 \begin{split}
\label{eq:LOMix}
\pi^0 &\rightarrow \pi^0 + \theta_{a\pi} a \approx \pi^0 + \epsilon \, \frac{K_{a\pi}\,m_a^2}{m_a^2 - m_\pi^2} a\,,\\
\eta &\rightarrow \eta + \theta_{a\eta}  a \approx \eta + \epsilon \, \frac{K_{a\eta}\,m_a^2 + m_{a\eta}^2}{m_{a}^2 - m_\eta^2} a\,,\\
\eta' &\rightarrow \eta' + \theta_{a\eta'}  a \approx \eta' + \epsilon \, \frac{K_{a\eta'}\,m_a^2 + m_{a\eta'}^2}{m_{a}^2 - m_{\eta'}^2} a\,,\\
\end{split}
 \end{align}
where $\epsilon = F_0/\Lambda$ and small mixing angles are assumed, i.e.~the ALP mass should not be too close to any of the pseudoscalar masses.
The kinetic mixing contributions $K_{aP}$ are given by
\begin{align}
 \begin{split}
  \label{eq:KinMixDef}
  K_{a\pi} &= 16 \pi^2 C_{GG} (\kappa_u - \kappa_d)\,,\\
  K_{a\eta} &= 16 \pi^2 C_{GG} \left( (\kappa_u + \kappa_d) \left(\frac{1}{\sqrt{3}}c(\theta) -\sqrt{\frac{2}{3}} s(\theta) \right) - \kappa_s \left(\frac{2}{\sqrt{3}} c(\theta) + \sqrt{\frac{2}{3}} s(\theta) \right) \right)\,,\\
    K_{a\eta'} &=16 \pi^2 C_{GG} \left( (\kappa_u + \kappa_d) \left(\frac{1}{\sqrt{3}}s(\theta) +\sqrt{\frac{2}{3}} c(\theta) \right) - \kappa_s \left(\frac{2}{\sqrt{3}} s(\theta) - \sqrt{\frac{2}{3}} c(\theta) \right)  \right)\,,\\
    \end{split}
 \end{align}
while the mass mixing terms $m_{aP}$ are
\begin{align}
 \begin{split}
   \label{eq:MassMixDef}
  m_{a\eta}^2 &= 16 \pi^2 C_{GG}\,\frac{2\,\sqrt{6}\, B_0 }{\text{Tr}[M_q^{-1}]}\,s(\theta)\,,\\
  m_{a\eta'}^2 &= -16 \pi^2 C_{GG}\,\frac{2\,\sqrt{6}\,B_0 }{\text{Tr}[M_q^{-1}]}\,c(\theta)\,,
  \end{split}
   \end{align}
where $B_0 = m_{\pi}^2 / (m_u + m_d)$ and $c(\theta)\equiv \cos(\theta)$ as well as $s(\theta)\equiv \sin(\theta)$ describe the $\eta\text{--}\eta'$ mixing.\footnote{We choose $\theta = -13^\circ$~\cite{Bossi:2008aa} yielding $m_\eta \approx 537\,\mathrm{MeV}$ and $m_{\eta'} \approx 1.15\,\mathrm{GeV}$ which is of sufficient accuracy for our purposes.}

\subsection{Interactions of light ALPs with photons and nucleons}
\label{subsec:ALPphotnucInt}
The coupling studied most in terms of constraints and experimental prospects in the ALP literature is the one to photons, see e.g.~refs.~\cite{Dolan:2017osp,Jaeckel:2017tud,Lee:2018lcj,Cadamuro:2011fd,Jaeckel:2010ni,Hewett:2012ns,Mimasu:2014nea,Dobrich:2019dxc}. It is therefore crucial to capture all of the sizeable contributions to the photon coupling in our EFT set-up. Here it is important to notice that the tree-level coupling in eq.~\eqref{eq:EWCoupl} generated after EWSB receives further hadronic contributions from the gluonic ALP coupling as is evident from eq.~\eqref{eq:LagrChirRot}. We therefore define an effective ALP-photon coupling~\cite{Bauer:2017ris,Domingo:2016yih,PhysRevD.98.030001}
\begin{align}
 \begin{split}
 \label{eq:EffPhotCoupl}
C^\text{eff}_{\gamma\gamma} \approx&\, C_{\gamma\gamma} - 1.92\, C_{GG} + \frac{1}{16\pi^2} K_{a\pi} \frac{m_a^2}{m_a^2 - m_\pi^2 }  + \frac{1}{16\pi^2} \frac{K_{a \eta} m_a^2 + m_{a \eta}^2}{m_a^2 - m_\eta^2 }\\
&+ \frac{13.6}{10.8} \frac{1}{16\pi^2} \frac{K_{a {\eta'}} m_a^2 + m_{a \eta'}^2}{m_a^2 - m_{\eta'}^2 }\,,
\end{split}
   \end{align}
where the precise coefficient multiplying $C_{GG}$ can be obtained by including higher orders in the computation~\cite{diCortona:2015ldu}.\footnote{We neglect the small uncertainties of 0.04 in the determination of this coefficient for our discussion. We note however that recently ref.~\cite{Lu:2020rhp} computed the coefficient to be 2.05, which suggests that uncertainties may in fact be larger.} We have also taken the mixing induced contributions into account, which are valid for ALP masses not too close to the respective pseudoscalar mass. We emphasise that the ALP-photon coupling is very generic in the sense that it receives a contribution from all the Wilson coefficients studied in this work and will therefore always be sizeable unless an accidental cancellation occurs. This simple observation demonstrates that studying one operator at a time, as usually done in the ALP literature, is insufficient to capture the full phenomenology of ALP models.

Using this effective photon coupling the decay width to photons can then be written as
\begin{align}
 \begin{split}
\label{eq:ALPdecaywidth}
 \Gamma_{a\gamma\gamma} &= \frac{4 \pi \alpha^2 m_a^3}{\Lambda^2} |C^\text{eff}_{\gamma\gamma}|^2\,,
 \end{split}
 \end{align}
where $\alpha$ is the electromagnetic fine-structure constant. We point out that, since we assume negligible tree-level interactions between ALPs and electrons, the ALP will decay dominantly into photons, i.e. $\Gamma^{\,a}_\text{tot} \approx \Gamma_{a\gamma\gamma}$, and hence its lifetime will be very sensitive to its mass. Loop-induced decays into electrons/muons are negligible~\cite{Gavela:2019wzg} and the hadronic decay channel $a \rightarrow 3 \pi$ is relevant only for $m_a > 3 m_{\pi}$, which is out of the mass range considered in this work.

For $C_{GG} \neq 0$ the ALP will also couple to nucleons. These interactions can be determined by including protons and neutrons into the chiral EFT. Defining the nucleon coupling by $\mathcal{L} = \partial_\mu a / (2 \Lambda)\,C_N \,\bar{N} \gamma^\mu \gamma_5 N $, where $N$ is a nucleon field, we obtain\footnote{More precise calculations have been performed in~\cite{diCortona:2015ldu} and most recently in~\cite{Vonk:2020zfh}. The values used here are also consistent with their results. Note that for the comparison one has to identify $1/f_a = -32 \pi^2 C_{GG}/\Lambda$.}~\cite{DiLuzio:2017ogq}
\begin{align}
 \begin{split}
\label{eq:NucGluCoupl}
 C_p + C_n &\approx 16\pi^2 C_{GG}\,,\\
 C_p - C_n &\approx \frac{1.273}{3}\, 32\pi^2 C_{GG}\,.
 \end{split}
 \end{align}
Note that these relations imply that for an ALP coupling predominantly to gluons the effective proton coupling $C_p$ is about an order of magnitude larger than the effective neutron coupling $C_n$.

\subsection{\texorpdfstring{Computation of $K^+ \rightarrow \pi^+ + a$}{Computation of K to pi a}}
\label{subsec:KpiaComp}

One of the most sensitive laboratory probes of MeV-scale ALPs are searches for the rare decay of a charged kaon to a charged pion and an ALP, where the ALP escapes from the detector as an invisible particle~\cite{Gavela:2019wzg,Izaguirre:2016dfi}. Past experimental searches in this channel provide stringent constraints on ALP models~\cite{Adler:2004hp,Artamonov:2009sz}, while upcoming experiments offer substantial discovery potential. 

For $C_{WW} \neq 0$ the transition occurs via a loop-level FCNC, where the ALP is emitted from the internal W boson line of the penguin diagram. This transition generates an effective interaction~\cite{Izaguirre:2016dfi}
\begin{align}
 \mathcal{L} = - g_{sd} \,\partial_\mu a\,\, \bar{d} \gamma^\mu P_L s+ \text{h.c.}\,,
 \label{eq:transition}
\end{align}
with the effective coefficient
\begin{align}
\label{eq:FCNCeffCoeff}
 g_{sd} =& \frac{6\,G_F^2\, m_W^4}{\pi^2 } \frac{C_{WW}}{\Lambda} \sum_{q=c,t} V^*_{qd} V_{qs} f(m_q^2/m_W^2)\,.
\end{align}
Here we have neglected the up-quark contribution and introduced the loop function
\begin{align}
 f(x) = \frac{x (1+x(\ln(x) -1))}{(1-x)^2}\,.
\end{align}
With this effective interaction we can then determine the decay width to be
\begin{align}
 \Gamma(K^+ \rightarrow \pi^+ + A) = \frac{|g_{sd}|^2}{64\,\pi \,m^3_{K^+}}(m^2_{K^+} - m^2_{\pi^+})^2\, \lambda^{1/2}(m^2_{K^+},\,m^2_{\pi^+},\,m^2_a)\,,
\end{align}
where $\lambda(x,y,z) = x^2 + y^2 + z^2 - 2xy -2xz -2xy$. We have neglected the hadronic form factor as it is close to unity~\cite{Dolan:2014ska,Marciano:1996wy,Carrasco:2016kpy}.

In the case of gluonic ALP couplings, the ALP mixes with the pseudoscalar mesons as discussed in section~\ref{subsec:Mixing} and will therefore participate in the $K^+$ decay through mixing effects. Taking into account that $K^+ \rightarrow \pi^+ a$ involves a change of isospin by $1/2$, one expects the $\Delta I = 1/2$ rule to apply~\cite{Bardeen:1986yb}. By employing a chiral Lagrangian (see appendix \ref{app:KtoPiMix} for calculational details), we can then approximately relate the ALP amplitude to the amplitude for $K^0 \rightarrow \pi^+ \pi^-$:
\begin{align}
\label{eq:KtoPiMixAmpl}
  &\mathcal{M}(K^+ \rightarrow \pi^+ a) \approx \theta_\text{mix} \,\mathcal{M}(K^0 \rightarrow \pi^+ \pi^-)\\
  \notag
  &= \Big[\theta_{a\pi} \frac{3A_2}{2A_0} e^{i(\chi_2 - \chi_0)}  + \theta_{a\eta} \sqrt{\frac{2}{3}} \big(c(\theta) - \sqrt{2} s(\theta)\big) + \theta_{a\eta'} \sqrt{\frac{2}{3}} \big(\sqrt{2} c(\theta) + s(\theta)\big)\Big] \mathcal{M}(K^0 \rightarrow \pi^+ \pi^-)\,.
\end{align}
Here $A_i$ and $\chi_i$ describe the amplitudes and phases in the isospin decomposition of the kaon amplitudes according to the notation in refs.~\cite{Cirigliano:2003gt,Cirigliano:2011ny}. The first term corresponds to the contribution of the $ K^+ \rightarrow \pi^+ \pi^0$ decay, which is usually omitted because of the $A_2 / A_0 \approx 1/22$ suppression. However, for $m_a \approx \mathcal{O}(100\,\mathrm{MeV})$ one finds $\theta_{a\pi} \gg \theta_{a\eta}, \theta_{a\eta'}$, which can overcome this suppression. The resulting decay width reads~\cite{Bardeen:1986yb,Alves:2017avw}
\begin{align}
\label{eq:K2piaMix}
 \text{BR}(K^+ \rightarrow \pi^+ a) \approx \lvert \theta_\text{mix} \rvert^2\, \frac{\tau_{K^+}}{\tau_{K_S^0}}\,  \text{BR}(K^0_S \rightarrow \pi^+ \pi^-) \frac{\lvert\vec{p}_a\rvert}{\lvert\vec{p}_\pi\rvert} D_{\pi\pi}^2\,,
\end{align}
where $\tau$ are the kaon lifetimes, $D^2_{\pi\pi} \approx 1/3$ corrects for the absence of final-state interactions for $\pi^+ a$~\cite{Truong:1980pf,Antoniadis:1981zw} and the three-momenta are $\lvert\vec{p}_X\rvert = \lambda^{1/2}(m_K^2,\,m_\pi^2,\,m_X^2)/(2m_K)$ in the lab frame. We emphasize that the computation presented above is only of approximate nature and is subject to uncertainties such as the $\eta\text{--}\eta'$ mixing that can not be described accurately at leading order in a chiral EFT. We refer to ref.~\cite{Alves:2017avw} for a more detailed discussion.

We note that ALP-quark couplings would in general also give a contribution to the process $K^+ \to \pi^+ a$. The resulting effect is however not easy to calculate in an EFT approach, because the loop-induced FCNC $s\rightarrow d a$ is divergent~\cite{Batell:2009jf}. In other words, it becomes difficult to reliably calculate experimental bounds and discovery prospects in the $K^+ \rightarrow \pi^+ a$ channel without a specific underlying model. To obtain a rough estimate for the case of flavour-universal couplings, one can replace the UV divergence by a leading logarithm~\cite{Batell:2009jf}. Adopting this prescription we find that for an ALP that couples dominantly to quarks the most interesting regions of parameter space are already excluded by existing constraints. We will therefore not discuss this case in more detail and focus instead on the interactions with gauge bosons.

\section{Experimental and observational constraints}
\label{sec:ExpConstr}

In the following we will discuss the most important bounds from fixed-target experiments, collider searches and astrophysics. We will, however, not consider cosmology constraints as they are very model-dependent, see ref.~\cite{Depta:2020wmr} for a detailed recent discussion. Instead we will focus on the general aspects of a light pseudoscalar particle interacting dominantly with SM gauge bosons independent of its precise cosmological history.
\subsection{\texorpdfstring{$K^+ \rightarrow \pi^+ + \text{inv}$}{K to Pi invisible}}
The fixed-target experiment E787 and its successor E949 have studied the decays of charged kaons at rest into a single charged pion and have placed constraints at 90\% confidence level (C.L.) on the process $K^+ \rightarrow \pi^+  + X^0$~\cite{Adler:2004hp,Artamonov:2009sz}, where $X^0$ denotes a new, long-lived particle. To interpret the experimental bounds on the branching ratio in our context, we have to ensure that the ALP in fact escapes from the detector before decaying. We do this by including a probability factor that the ALP does not decay inside of the decay volume
\begin{align}
 \begin{split}
 \label{eq:KpiAExpTheo}
  \text{BR}(K^+ \rightarrow \pi^+ a)^{\text{exp}} = \text{BR}(K^+ \rightarrow \pi^+ a)^{\text{theo}}\, e^{-L_\text{det}/l_\text{a}}\,.
  \end{split}
 \end{align}
Here $L_\text{det}$ describes the characteristic size of the detector and $l_{a} = \lvert\vec{p}_a\rvert/(m_a \Gamma^{\,a}_\text{tot})$ is the ALP decay length determined by its momentum $\lvert\vec{p}_a\rvert$ in the lab frame. For E787 and E949 we take \mbox{$L_\text{det} = 4\,\mathrm{m}$} and calculate $\lvert\vec{p}_a\rvert$ given that the kaons decayed at rest.\footnote{Note that E949 includes lifetime-dependent bounds on $K^+ \rightarrow \pi^+  + X^0$. These are approximately reproduced by the exponential probability factor, which is only of relevance for $m_a \gtrsim 150\,\mathrm{MeV}$. Below this mass the ALP is always sufficiently long lived for the coupling values of interest.}

We will also include prospects for NA62, a presently running experiment aiming to measure the ultra-rare \mbox{$K^+ \rightarrow \pi^+ \nu \bar{\nu}$} channel, which has an SM prediction of $\text{BR}(K^+ \rightarrow \pi^+ \nu \bar{\nu}) = (8.4 \pm 1.0)\cdot 10^{-11}$~\cite{Buras:2006gb,Brod:2010hi,Buras:2015qea}, with $10\%$ precision~\cite{NA62:2017rwk}. In this channel the $\nu \bar{\nu}$ pair escapes the detector without a trace, giving rise to a missing mass $m_\text{miss}^2 = (p_K - p_\pi)^2$, where $p_K$ and $p_\pi$ are the kaon and pion four-momenta. When reinterpreting this search for $K^+ \rightarrow \pi^+ a$ we need to account for the fact that the SM process has a 3-particle final state. In this case the missing mass distribution is spread kinematically whereas for the 2-particle final state $m_\text{miss}^2$ will be peaked. Note that, while NA62 cannot search for ALPs with mass $m_a \approx m_\pi$ due to prohibitively large backgrounds from $K^+ \to \pi^+ \pi^0$, it is sensitive to both the case $m_a < m_\pi$ (signal region 1) and $m_a > m_\pi$ (signal region 2).

\begin{figure}[t]
\centering
\includegraphics[width=6.9cm]{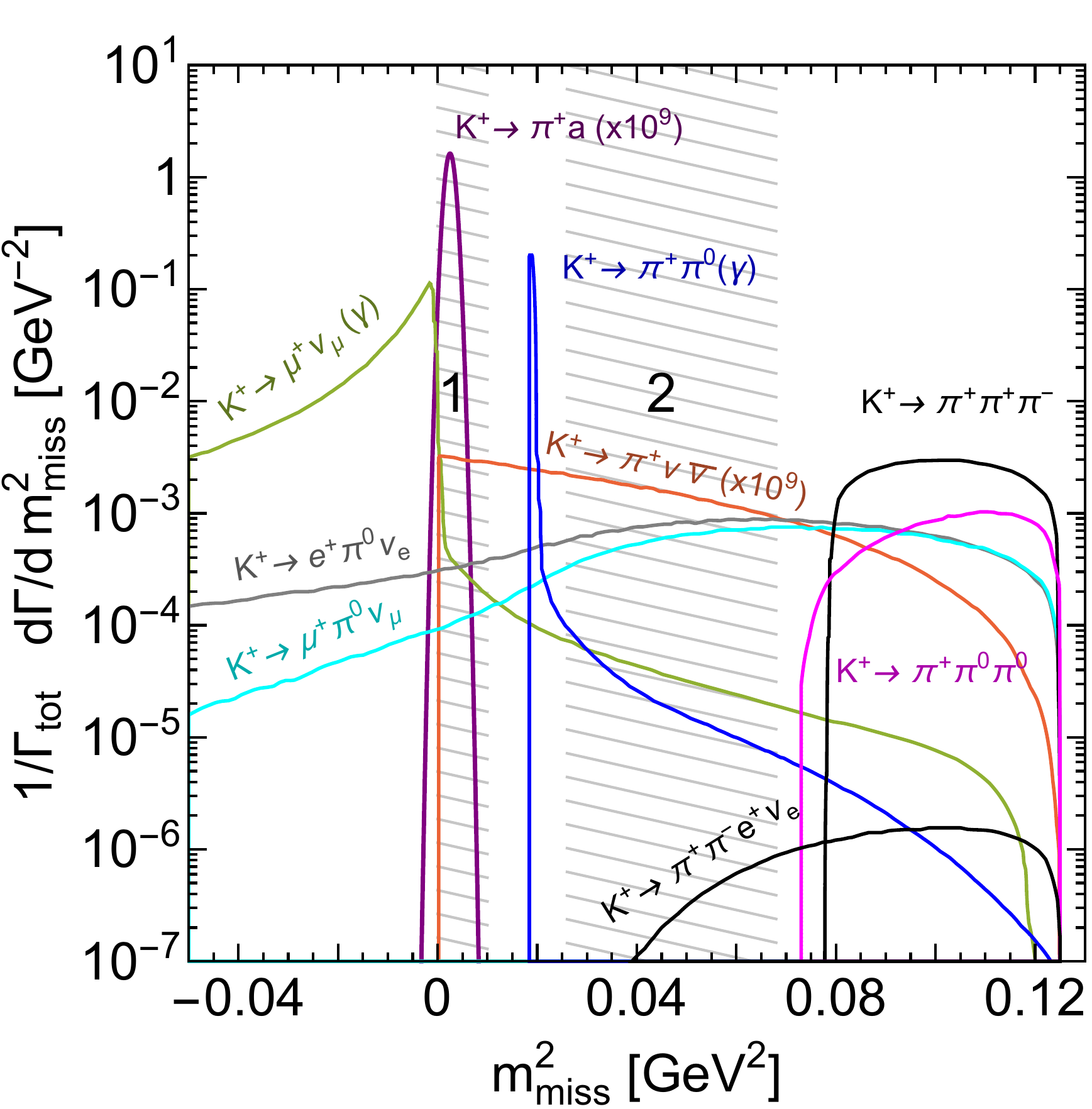}
\caption{Differential distribution of $m_\text{miss}^2 $ for different $K^+$ decays taken from refs.~\cite{NA62:2017rwk,CortinaGil:2018fkc} and supplemented by the prediction for $K^+ \rightarrow \pi^+ a $ (purple) with $C_{WW}/\Lambda = 5 \cdot 10^{-4}\,\mathrm{TeV}^{-1}$ and $m_a = 50\,\mathrm{MeV}$, assuming a $m_\text{miss}^2$ resolution of $0.001\,\mathrm{GeV^2}$. Here the regions labelled as ``1'' and ``2'' refer to the two signal regions of NA62, which are separated by the $K^+ \rightarrow \pi^+ \pi^0$ background. 
}\label{fig:MissMassDistr}
\end{figure}
As an example we plot the missing mass distribution $m_\text{miss}^2$ in figure~\ref{fig:MissMassDistr} assuming $C_{WW}/\Lambda = 5 \cdot 10^{-4}\,\mathrm{TeV}^{-1}$ and $m_a = 50\,\mathrm{MeV}$ corresponding to $\text{BR}(K^+ \rightarrow \pi^+ a) \approx 4.1 \cdot 10^{-12}$. For this particular choice of mass this signal falls into the signal region 1 of NA62. One notices that despite $K^+ \rightarrow \pi^+ a$ having a much smaller branching ratio than $K^+ \rightarrow \pi^+ \nu \bar{\nu}$ for these parameter points, the different kinematics imply that this ALP channel would have a visible impact on the measured distribution. However, as there are no official projections provided by NA62 concerning their sensitivity for the $K^+ \rightarrow \pi^+ a$ channel, we assume that NA62 will improve the experimental bounds by E787 and E949 by roughly an order of magnitude based on the information given in table 2 of ref.~\cite{Fantechi:2014hqa}.\footnote{We also point out that the experiment KLEVER~\cite{Ambrosino:2019qvz}, currently in the process of being designed to measure $\text{BR}(K^0_L \rightarrow \pi^0 \nu \bar{\nu})$ with better than 20\% accuracy, could even exceed the reach of NA62 assumed here~\cite{Beacham:2019nyx}.} For eq.~\eqref{eq:KpiAExpTheo} we take $\lvert\vec{p}_a\rvert = E_a = 37\,\mathrm{GeV}$ and $L_\text{det} = 200\,\mathrm{m}$ in the case of NA62~\cite{Kitahara:2019lws}.

\subsection{Beam dump experiments}
\label{subsec:ConstrBDExp}

In beam dump facilities ALPs can be produced not only in flavour changing meson decays, but also via their direct couplings to photons and gluons, e.g.\ via the Primakoff process. Important constraints come from the E137 experiment at SLAC~\cite{Bjorken:1988as}, where a search for ALPs with dominant photon couplings was performed. The original analysis was extended to the full ALP parameter space in ref.~\cite{Dolan:2017osp}, which uses the null result to obtain a bound at 95\% C.L.\,. The same constraint also applies in the case of a dominant ALP-gluon coupling, for which the induced photon coupling (see  eq.~\eqref{eq:EffPhotCoupl}) dominates the production in electron beam dump experiments.

Proton beam dump experiments~\cite{Bergsma:1985qz,Blumlein:1990ay,Blumlein:1991xh} also offer an additional promising search strategy for ALPs with gluon couplings. In refs.~\cite{Ariga:2018uku,Beacham:2019nyx} the constraints by CHARM have been revisited for the case that the production occurs through mixing with pseudoscalar mesons. The resulting production is found to be strongly enhanced compared to the case of couplings only to photons for ALPs with mass $m_a \gtrsim 100\,\mathrm{MeV}$. Moreover, due to their higher beam energy, proton beam dump experiments are particularly sensitive to relatively large ALP masses and couplings, for which ALPs produced in an electron beam dump would decay before reaching the detector.
However, the information provided in refs.~\cite{Ariga:2018uku,Beacham:2019nyx} is insufficient to allow for a reinterpretation of these bounds for ALPs with additional couplings to photons or under the assumption of different $\eta$ and $\eta'$ mixing angles. We will therefore not include this bound in the following but stress that the sensitivity of proton beam dump experiments to ALPs with several couplings and large masses would be very interesting to explore in a proper simulation (see refs.~\cite{Dobrich:2019dxc,Dobrich:2018jyi}).

\subsection{\texorpdfstring{$Z \rightarrow \gamma + \text{inv.}$}{Z to gamma inv}}

In section~\ref{subsec:ALPEFT} it was shown that the various ALP interactions in the electroweak sector are highly correlated and are therefore expected to arise simultaneously. In addition to the $W^\pm$ boson and photon coupling, the coefficient $C_{\gamma Z}$ can be relevant as it induces the $Z \rightarrow \gamma + a$ decay. The decay width for this process reads~\cite{Dror:2017nsg, Bauer:2017ris}
\begin{align}
 \begin{split}
  \Gamma(Z \rightarrow \gamma a) = \frac{8 \pi \alpha \alpha(m_Z) m_Z^3}{3 s_w^2 c_w^2} \left(\frac{|C_{\gamma Z}|}{\Lambda}\right)^2 \left(1 - \frac{m_a^2}{m_Z^2} \right)^3\,.
  \end{split}
 \end{align}
A search in this channel was performed by L3 at LEP at the $Z$ resonance. The null-results were converted into a bound on the branching ratio  as a function of the photon energy, which for small $m_a$ corresponds to $\text{BR}(Z \rightarrow \gamma a) \lesssim 10^{-6}$ at 95\% C.L.~\cite{Dolan:2017osp,Acciarri:1997im}.

\subsection{Astrophysical bounds}
\label{subsec:AstroBounds}
For MeV-scale ALPs that couple to photons and nucleons astrophysical probes also provide relevant constraints. In our discussion cooling constraints and the missing ``ALP burst'' from SN1987A as well as modifications of the stellar evolution of horizontal branch (HB) stars are relevant.

\subsubsection*{Cooling constraints from SN1987A}

In the standard scenario of a core-collapse supernova most of its energy is released through neutrino emission, constituting the main cooling mechanism for the proto-neutron star, which forms as a result of the collapse. However, if additional, new weakly interacting particles exist with masses below $\mathcal{O}(100\,\mathrm{MeV})$ these can also be produced in the SN core, which has a temperature of several tens of MeV. If this new energy loss mechanism is as efficient as (or even more efficient than) the energy loss through neutrinos, this would be in tension with the observed duration of the neutrino signal on Earth for the case of SN1987A. To make precise statements about the influence of a specific ALP on the supernova evolution one would in principle need to include all relevant effects in a detailed supernova simulation. Here we will instead use the approximate analytic criterion proposed by Raffelt~\cite{Raffelt:1996wa} stating that the energy loss due to new particles should not exceed the neutrino luminosity $L_{\nu} = 3\cdot 10^{52}\,\text{erg}/\text{s}$.\footnote{Note that recently ref.~\cite{Bar:2019ifz} critically assessed the standard picture of a core-collapse for SN1987A assumed here.}

In order to determine the luminosity $L_a$ of ALPs produced in a proto-neutron star, we employ the ansatz from refs.~\cite{Chang:2016ntp,Chang:2018rso} and extend it to ALP-photon couplings. The luminosity is defined by a volume integral over the volume emission rate of ALPs within the neutrinosphere $R_\nu$, beyond which neutrinos stream freely, as ALPs produced inside and escaping this region pose a new energy loss mechanism. However, for large ALP couplings the interaction will be so strong that most ALPs produced in this region do not actually escape but get absorbed again, which is referred to as the ``trapping regime''. This is taken into account by including an optical depth factor in the integration characterizing the probability that an ALP produced within the neutrinosphere reaches the radius $R_\text{far}$, beyond which the ALP energy cannot get reprocessed efficiently into neutrino energy~\cite{Chang:2016ntp}. We provide more details on the luminosity formula and the ALP energy distribution contributing to the luminosity in appendix~\ref{app:LumDetails}.

For ALPs coupling dominantly to photons Primakoff conversion involving protons, $\gamma + p \rightarrow a + p$, provides the main production mechanism~\cite{Raffelt:1985nk}. In the case of ALP-gluon couplings bremsstrahlung in nucleon scattering $N + N \rightarrow N + N + a$ is most relevant~\cite{Raffelt:1996wa,Raffelt:1990yz}. Processes involving electrons are negligible as their phase space is Pauli-blocked in a supernova core~\cite{Payez:2014xsa}. Combining all of these points, we obtain for the ALP luminosity
\begin{align}
 \begin{split}
  \label{eq:LumFormula}
  \hspace{-0.5cm}L_a &= \int_{r \leq R_\nu} \text{d}V \left(\int \frac{\text{d}^{3}\textbf{p}_a}{(2\pi)^3} \omega  \Gamma_a e^{-\omega/T}\,\beta\, e^{-\tau}+ \int \frac{2\text{d}^{3} \textbf{p}_\gamma}{(2\pi)^3} \frac{\omega \Gamma_{\gamma \rightarrow a}}{e^{\omega/T}-1} \beta\,e^{-\tau}\right)\,,\\
  &= \int_{r \leq R_\nu} \text{d}V \int_{m_a}^\infty \frac{\text{d}\omega}{2 \pi^2} \left(\omega^2  \,\Gamma_a e^{-\omega/T}\sqrt{\omega^2 - m_a^2} + 2 \frac{\omega^3 \Gamma_{\gamma \rightarrow a}}{e^{\omega/T}-1}\right) \beta \,e^{-\tau}\,.
  \end{split}
 \end{align}
The integration variables $\textbf{p}_a$ and $\textbf{p}_\gamma$ describe the ALP and photon momenta, $\Gamma_{\gamma \rightarrow a}$ and $\Gamma_{a}$ the production rates through Primakoff and bremsstrahlung, $T$ the temperature as a function of radius $r$ and $\tau \equiv \tau(m_a,\,\omega,\, r ,\, R_\text{far})$ is the optical depth. Following ref.~\cite{Chang:2018rso}, we set $R_\text{far} = 100\,\mathrm{km}$. For the Primakoff process we have used that the photon energy $\omega_\gamma$ is equivalent to the ALP energy $\omega_a \equiv \omega$ as the proton mass is much larger than all other energies involved~\cite{Raffelt:1985nk,Cadamuro:2011fd}. In the integration we have also taken into account that the ALP/initial photon energy has to be at least $m_a$ to produce an ALP of this mass. The effects of sizeable ALP masses, i.e. $m_a > 1\,\mathrm{MeV}$, are estimated by a phase space factor $\beta = \sqrt{1 - m_a^2/\omega^2}$~\cite{Chang:2018rso,Lee:2018lcj}. We now go into more detail concerning the different quantities entering eq.~\eqref{eq:LumFormula}:
\begin{itemize}
\item We use the axion absorptive width $\Gamma_a$ for nucleon brems\-strah\-lung given in eqs.~(4.2) and (4.3) in ref.~\cite{Chang:2018rso}, where we perform the replacement $f_a \rightarrow \Lambda$. For the one-pion exchange correction factor $\gamma_h$ we use the parameter values for an electron fraction of $Y_e = 0.1$ given in table 1 in ref.~\cite{Bartl:2016iok}. We stress, however, that the different parameter sets are very similar at higher densities, where most of the ALP production and capturing will occur. For the factor $\gamma_p$, which corrects for the finite pion mass and nucleon degeneracy, we take into account the fraction of nuclei that actually participate in the scattering. For example, an ALP with gluon couplings interacts much more strongly with protons than with neutrons. In this case we use a density of $Y_p\,\rho$ with the proton fraction $Y_p = 0.3$ as input parameter.
\item The transition rate of a (massless) photon with energy $\omega$ into an ALP of the same energy via the Primakoff process is given by~\cite{Raffelt:2006cw}
\begin{align}
 \begin{split}
\Gamma_{\gamma\rightarrow a} = \left(\frac{16 \pi \alpha C^\text{eff}_{\gamma\gamma}}{\Lambda}\right)^2  \frac{T \kappa^2}{32 \pi}\left[\left(1 + \frac{\kappa^2}{4 \omega^2} \right) \ln\left( 1 + \frac{4 \omega^2}{\kappa^2} \right) -1 \right]\,,
 \end{split}
 \end{align}
where $\kappa$ is the screening scale as the Coulomb potential only has a finite range in a plasma~\cite{Raffelt:1985nk}. Note that this transition rate is averaged over the photon polarizations, which is the origin of the factor $2$ appearing in eq.~\eqref{eq:LumFormula}. Taking into account that the protons are partially degenerate, the screening scale reads~\cite{Payez:2014xsa}
\begin{align}
 \begin{split}
 \kappa^2 = \frac{4 \pi \alpha}{T} n^{\text{eff}}_{p}\,,
 \end{split}
 \end{align}
with the effective number of proton targets $n^{\text{eff}}_{p}$. For simplicity we assume that $n^{\text{eff}}_{p} = 0.5\,n_p = 0.5\,Y_p\, \rho/m_p$, where $m_p$ is the proton mass, within the core region, i.e. $r \leq 10\,\mathrm{km}$, and $n^{\text{eff}}_{p} = n_p$ outside. This degeneracy has, however, only slight effects.
\item The optical depth includes contributions from inverse Primakoff processes, ALP decays into photons and inverse nucleon bremsstrahlung. Combining these, it reads (see appendix \ref{app:LumDetails} for the general definition)
\begin{align}
\label{eq:OptDepth}
 \begin{split}
 \hspace{-0.9cm}\tau =& (R_\text{far} - R_{\nu}) \beta^{-1}\left(2 \Gamma_{\gamma \rightarrow a}(R_\nu) + \gamma^{-1} \Gamma_{a\gamma\gamma} + \Gamma_{a}(R_\nu)\right)\\
 &+ \beta^{-1}\int_{r}^{R_\nu} dr' \left(2 \Gamma_{\gamma \rightarrow a}+ \gamma^{-1} \Gamma_{a\gamma\gamma}  + \Gamma_{a}\right)\,,
 \end{split}
\end{align}
where we included a factor 2 due to the two photon polarizations and a Lorentz factor $\gamma = \omega/m_a$~\cite{Lee:2018lcj,Raffelt:1988rx}. We also split the integration into $r<R_\nu$ and $r>R_\nu$ as the temperature and density profiles used do not go that far~\cite{Chang:2016ntp}. This will also give a conservative estimate of the effects in the outer region of the star as the temperature and density profile are kept constant beyond $r=R_\nu$ instead of letting them decrease.
\item The main temperature and density profile used in this work is the fiducial one as given in ref.~\cite{Chang:2016ntp}. We have also checked the exclusions contours for the profiles by Fischer ($11\,M_{\odot}$ and $18\,M_{\odot}$)~\cite{Fischer:2016cyd} as well as Nakazato ($13\,M_{\odot}$)~\cite{Nakazato:2012qf} with the parameters as given in table 2 of ref.~\cite{Chang:2016ntp} and have found the fiducial one to give the most conservative bound overall (see appendix~\ref{app:SNUncer} for more details). 
\end{itemize}
\begin{figure}[t]
\centering
\hspace{-0.4cm}
\includegraphics[width=6.9cm]{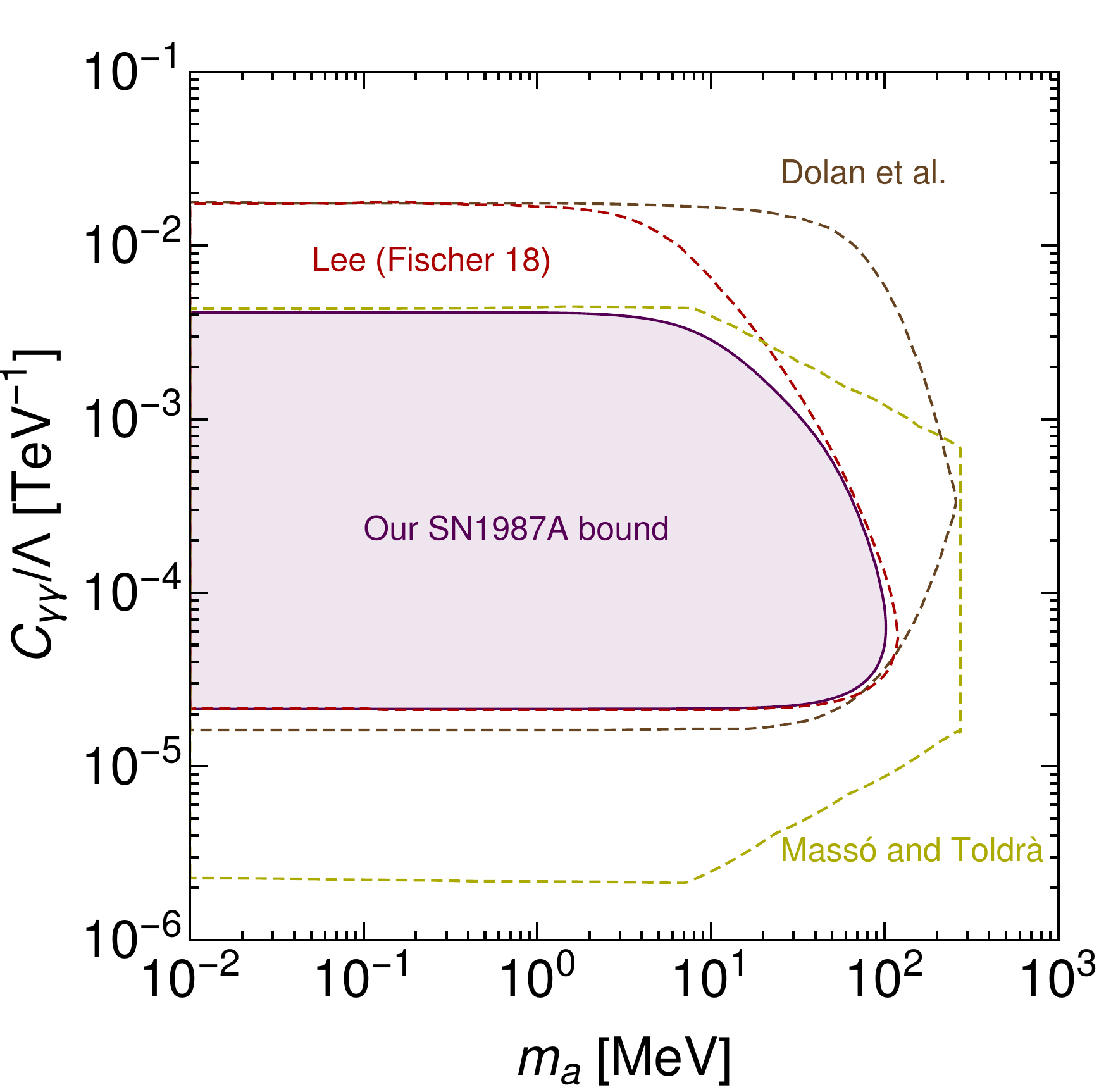}
\hspace{1cm}
\includegraphics[width=6.9cm]{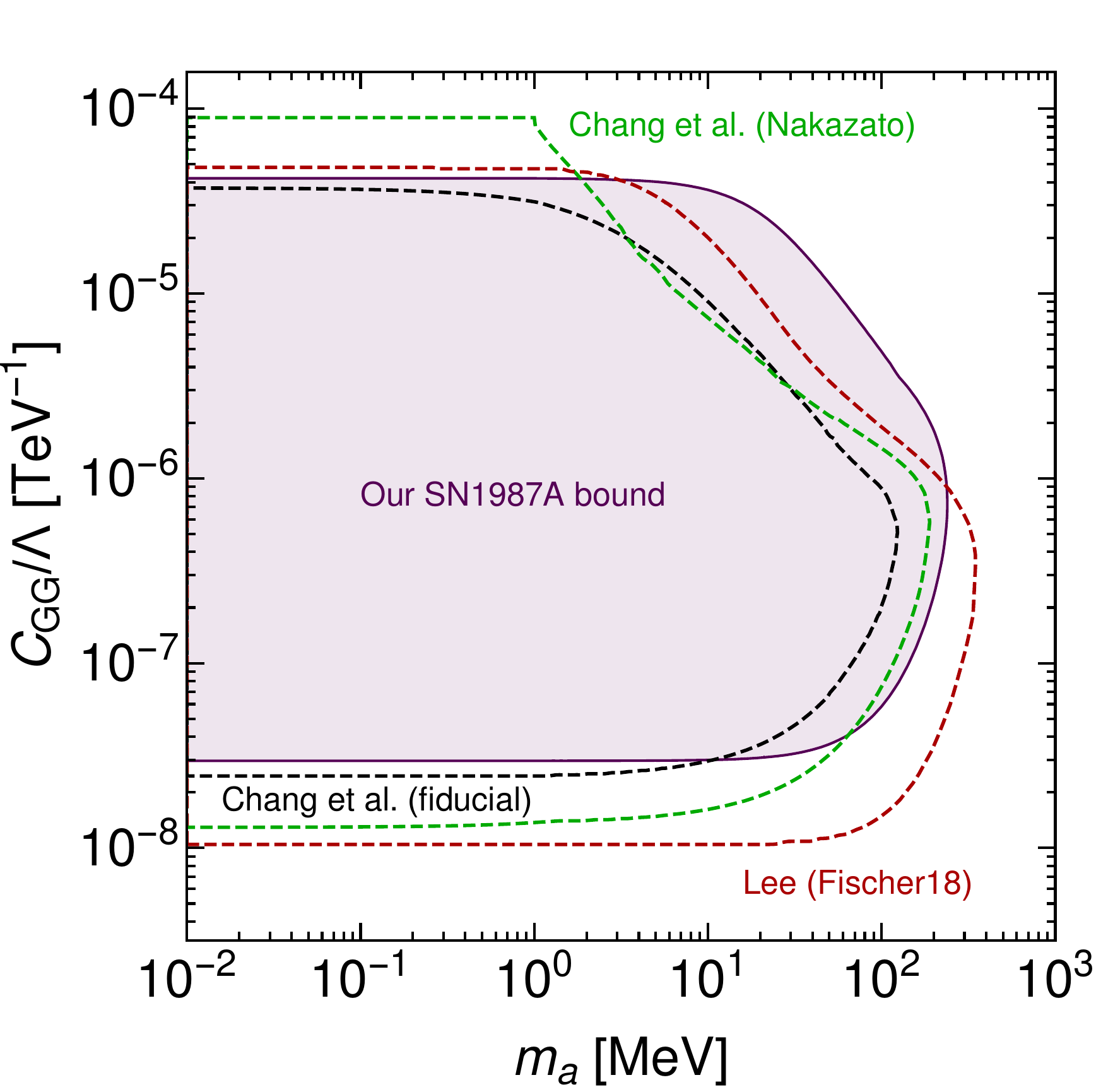}
\hspace{-0.1cm}
\caption{Comparison of the different supernova bounds for a dominant ALP-photon coupling (left) and a dominant ALP-gluon interaction (right). We have taken the brown curve from ref.~\cite{Dolan:2017osp}, the red curve from ref.~\cite{Lee:2018lcj}, the yellow one from ref.~\cite{Bauer:2017ris} (originally computed in ref.~\cite{Masso:1995tw}), the black as well as the green one from ref.~\cite{Chang:2018rso} and the purple coloured area corresponds to our result for the fiducial profile.
}\label{fig:ComparisonPhotSN}
\end{figure}

Let us now compare our results with those from the literature. It is important to stress that even when using the same cooling criterion, differences in the calculational approaches and the assumed temperature and density profiles typically lead to sizeable differences in the resulting bounds, which demonstrates the difficulty of extracting reliable supernova constraints. In the left panel of figure~\ref{fig:ComparisonPhotSN} we consider dominant ALP-photon couplings. We observe that the lower boundary, where ALP production becomes efficient enough to alter the duration of the neutrino signal (the free-streaming regime), is similar to the ones found in other recent calculations (Dolan et al.~\cite{Dolan:2017osp} and Lee~\cite{Lee:2018lcj}). In the trapping regime, however, for which we calculate a detailed optical depth factor, we obtain a weaker constraint.

In the right panel of figure~\ref{fig:ComparisonPhotSN}, we focus on the case of a dominant ALP-gluon coupling, for which bremsstrahlung production dominates over Primakoff processes. For the comparison we identify $g^2_{aNN}$ from ref.~\cite{Lee:2018lcj} and $1/(2f_a)^2$ from ref.~\cite{Chang:2018rso} with $(C_p^2 Y_p + C_n^2 Y_n)/\Lambda^2$. While our determination of the trapping regime for small ALP masses gives similar results in this case, our bound differs for larger ones as well as in the free-streaming regime. Most notably, we find that for the fiducial profile the supernova bound is essentially independent of the ALP mass for $m_a \lesssim 10\,\mathrm{MeV}$, consistent with the fact that most ALPs produced in the supernova have an energy well above 10 MeV (see appendix~\ref{app:LumDetails}). We also point out that ref.~\cite{Carenza:2019pxu} obtains a stronger bound on ALP-nucleon interactions for small ALP masses (in particular in the trapping regime), but does not consider ALPs with masses $m_a > 1\,\mathrm{MeV}$.

As mentioned before, assuming a different temperature and density profile and/or a smaller value for $R_\text{far}$ would lead to stronger bounds than those depicted in figure~\ref{fig:ComparisonPhotSN} (see appendix \ref{app:SNUncer} for more details). In these cases the bound on the ALP-photon coupling would be closer to the estimates in ref.~\cite{Lee:2018lcj} but the constraint on the ALP-gluon coupling would considerably exceed those derived there for the trapping regime. Given that the uncertainties affecting the SN cooling bounds are sizeable but difficult to quantify precisely, we adopt a conservative approach and use the bounds obtained with the fiducial profile as shown in figure~\ref{fig:ComparisonPhotSN} for the remainder of this work.

\subsubsection*{ALP burst from SN1987A}
In the coupling regime below the supernova cooling constraint ALPs are still produced within the supernova core, but they do not lead to significant energy loss. In ref.~\cite{Jaeckel:2017tud} it was pointed out that these ALPs are sufficiently long-lived to escape the supernova but would still decay before reaching Earth. This would have resulted in a (delayed) gamma ray burst that was, however, not observed. This missing ALP burst gives an important constraint for ALP-photon couplings below the cooling constraint.

\subsubsection*{Number count of events}
In addition to the two constraints discussed above, there is another bound on ALP-nucleon interactions often quoted in literature that excludes much more parameter space above the SN1987A cooling bound~\cite{Engel:1990zd}. In this strong coupling regime one has to consider that ALPs cannot escape from the core region but instead are emitted from the surface of a sphere~\cite{Burrows:1990pk}, the so-called axionsphere (analogous to the neutrinosphere). Although these ALPs do not lead to a sizeable energy loss, they can potentially reach the Earth and induce events in particle detectors. One therefore has to require that the number of events predicted in the Kamiokande experiment does not exceed observations~\cite{Engel:1990zd}. The ALPs that we consider, however, are too short-lived to reach the Earth so that this constraint does not apply.

What could play a role instead are ALPs produced on the axionsphere, leaving the supernova and then decaying to photons before reaching the Earth~\cite{Hall:2004qd}. The absence of a gamma ray signal could then lead to additional constraints similar to the ALP burst discussed before. However, in the coupling range of interest and for ALP masses above a few MeV the decay length is too short to reach the effective radius of $R_\text{eff} \approx 3 \cdot 10^{10}\,\mathrm{m}$ needed to leave the supernova~\cite{Jaeckel:2017tud,Kazanas:2014mca}. Of course, even in this case one would need to ensure that not too much energy is deposited in the mantle and envelope as most of the energy is liberated by neutrinos~\cite{Hall:2004qd,Raffelt:1999tx,Sung:2019xie}. A careful study of these issues is beyond the scope of this work.

\subsubsection*{HB stars}
The existence of light ALPs would also influence stellar evolution. Strongly affected by ALP-photon couplings are horizontal branch (HB) stars, i.e.\ stars that have entered the helium burning phase~\cite{Raffelt:1987yu}. An additional energy loss through ALPs leads to a faster contraction of the core region and hence an increase in temperature~\cite{Frieman:1987ui}. This then results in a faster burning of the helium fuel and reduces the lifetime of horizontal branch stars. This influence has been studied in the context of globular clusters, where one can determine the relative number of horizontal branch stars and red giants, which are not as affected by the Primakoff production of ALPs due to their degenerate cores~\cite{Raffelt:1998fy}. As the observed ratio is within 10\% of the prediction, new energy loss contributions can be constrained~\cite{Raffelt:2006cw,Raffelt:1996wa}. The resulting bounds on the ALP-photon coupling have been computed in a simplified manner in ref.~\cite{Cadamuro:2011fd}, which we will use in the following. Note that for ALPs that couple to gluons the induced ALP-photon interaction dominates over the ALP-nucleon interaction, because the resulting Compton cross section is strongly momentum suppressed~\cite{Raffelt:1996wa}.

\section{Results}
\label{sec:results}

In this section we combine the various constraints discussed above to determine the viable regions of parameter space and explore the prospects for probing ALPs with NA62. We consider first the case that ALPs couple only to electroweak gauge bosons, then explore the case of gluonic couplings and finally study the impact of varying all couplings simultaneously.

\subsection{Couplings to electroweak gauge bosons only}
\label{subsec:ResultEW}
\begin{figure}[t]
\centering
\includegraphics[width=6.75cm]{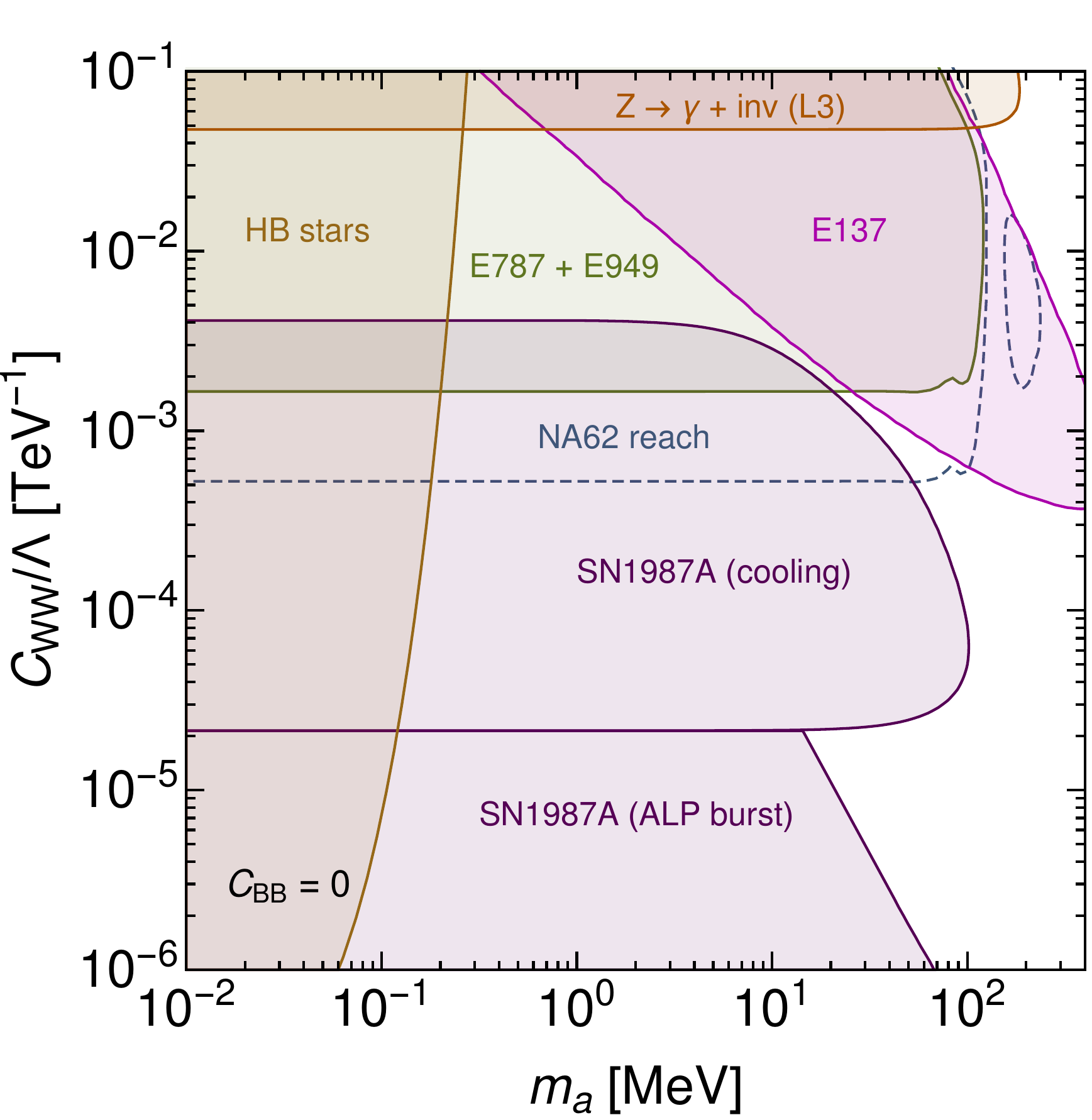}
\hspace{1cm}
\includegraphics[width=6.75cm]{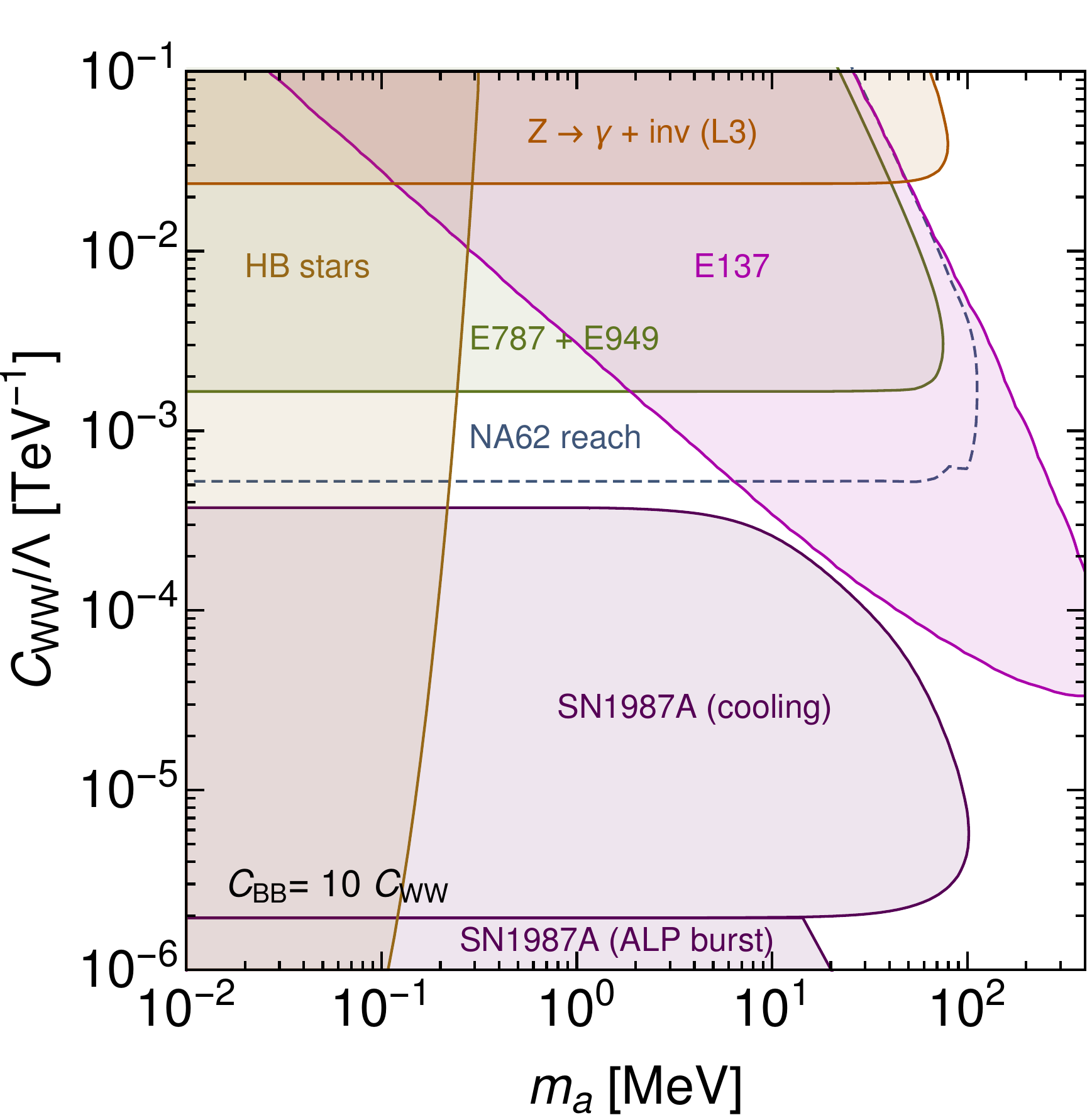}
\caption{Experimental constraints and prospects on ALPs interacting dominantly with electroweak gauge boson. On the left we focus on a sizeable value of $C_{WW}$. On the right we additionally assume $C_{BB} = 10\,C_{WW}$ as an example of how to circumvent the supernova bounds.}\label{fig:ConstraintsEW}
\end{figure}

Let us first consider the case of dominant couplings to $SU(2)_L$ gauge bosons, i.e.\ the case that all Wilson coefficients other than $C_{WW}$ can be neglected. The corresponding constraints are shown in the left panel of figure~\ref{fig:ConstraintsEW}. One finds that the combined constraints from HB stars, SN1987A, L3, E137 and E787/E949 exclude essentially all ALPs with mass $m_a \lesssim 100\,\mathrm{MeV}$ (unless $C_{WW} / \Lambda$ is much smaller than what is considered in this figure) and in particular the entire parameter region that can be probed with NA62. This conclusion relies crucially on the fact that couplings to $SU(2)_L$ gauge bosons induce both couplings to $W^\pm$ bosons (which lead to constraints on FCNCs from E787/E949) and couplings to $Z$ bosons and photons (which result in all the other constraints) after electroweak symmetry breaking. Previous studies have often found weaker constraints by either neglecting the ALP-photon coupling (as done e.g.\ in ref.~\cite{Gavela:2019wzg}) or assuming the presence of another light species that allows for ALPs to decay dominantly invisibly (see ref.~\cite{Izaguirre:2016dfi}).

However, it is also possible to relax the constraints without leaving the EFT framework introduced above. Since $C_{\gamma\gamma} = C_{WW} + C_{BB}$ it is possible to shift the constraints that depend on the ALP-photon coupling relative to the constraints that depend on the ALP-$W$ coupling by considering the case $C_{BB} \neq 0$. As an example, we consider in the right panel of figure~\ref{fig:ConstraintsEW} the case that $C_{BB} = 10 \, C_{WW}$, such that the value of $C_{\gamma\gamma}$ corresponding to a given value of $C_{WW}$ is substantially enhanced. The bounds that depend on the ALP-photon coupling are hence shifted downwards relative to the ones depending on FCNCs, so that the parameter space probed by NA62 falls into the trapping regime of SN1987A. Even when adopting a more agressive SN bound (see appendix \ref{app:SNUncer}), one can find an appropriate relation between $C_{BB}$ and $C_{WW}$ to open up parameter space for NA62 to probe as long as there is an unconstrained region between the bounds from HB stars, SN 1987A and E137. By considering simultaneously the ALP couplings to $SU(2)_L$ and hypercharge gauge bosons, it is hence possible to open up parameter space where MeV-scale ALPs may be discovered in the laboratory. 

We note that in the case that $C_{BB} = 10 \, C_{WW}$ there are no constraints from E787/E949 (and no sensitivity of NA62) for $m_a \gtrsim 100 \, \mathrm{MeV}$, because the enhanced photon coupling implies that ALPs in this mass range would no longer escape from the detector before decaying. Instead, they would contribute to the channel $K^+ \to \pi^+ \gamma \gamma$, where the sensitivity is reduced by large SM backgrounds and branching ratios as large as $\mathcal{O}(10^{-6})$ are allowed~\cite{Kitahara:2019lws,Ceccucci:2014oza}.

Finally, we point out that constraints could potentially be relaxed even further in the case of destructive interference, if $C_{WW} \approx - C_{BB}$ such that the ALP-photon coupling is suppressed. The constraints that depend on this coupling would then be shifted upwards relative to the constraints that depend on FCNCs. However, given the strength of the supernova constraints, a very precise cancellation would be required in order to evade these constraints entirely, making this solution less attractive.

\subsection{Couplings to all Standard Model gauge bosons}
\label{subsec:ResultEWG}
\begin{figure}[t]
\centering
\includegraphics[width=6.75cm]{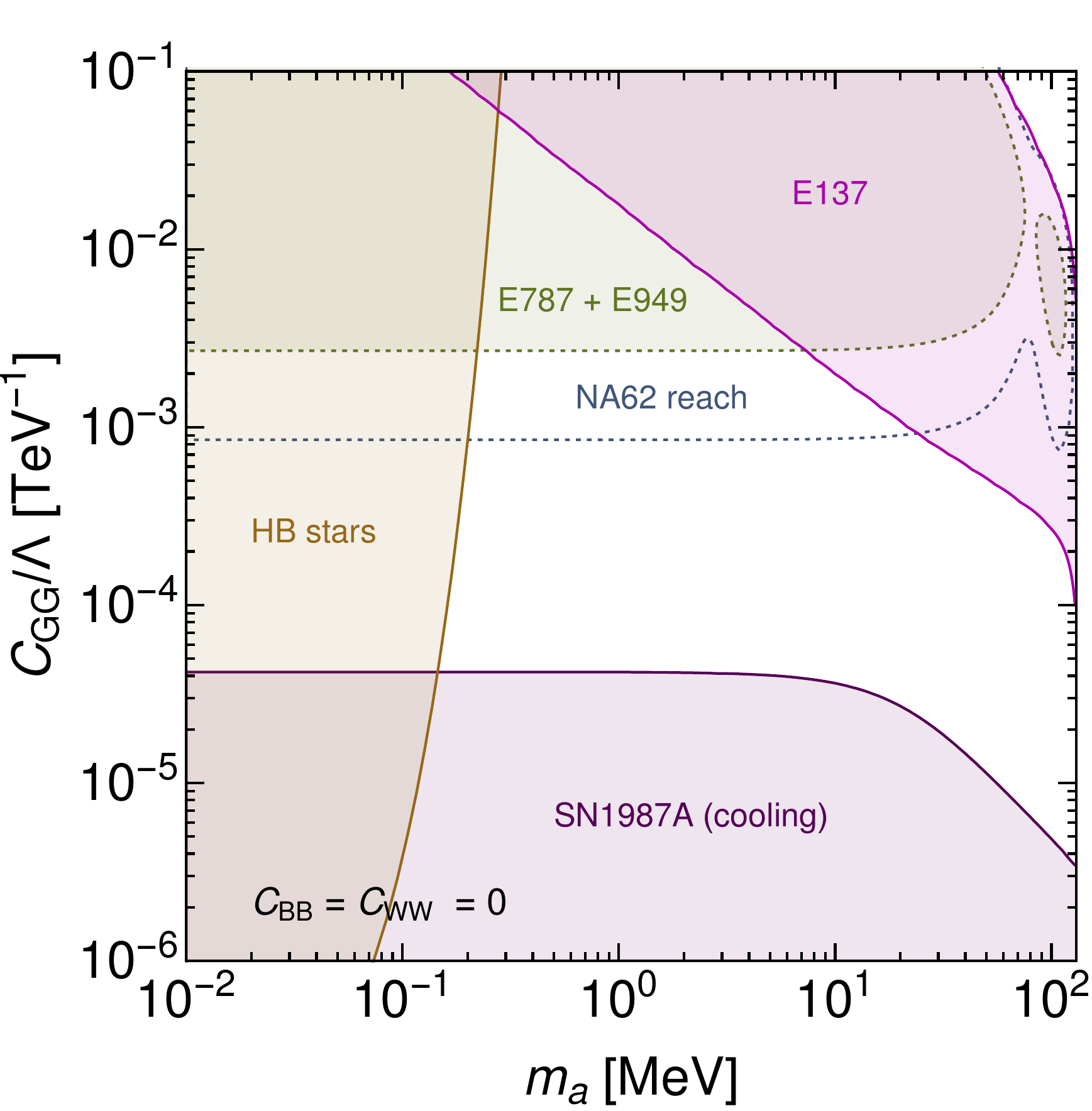}
\hspace{1cm}
\includegraphics[width=6.75cm]{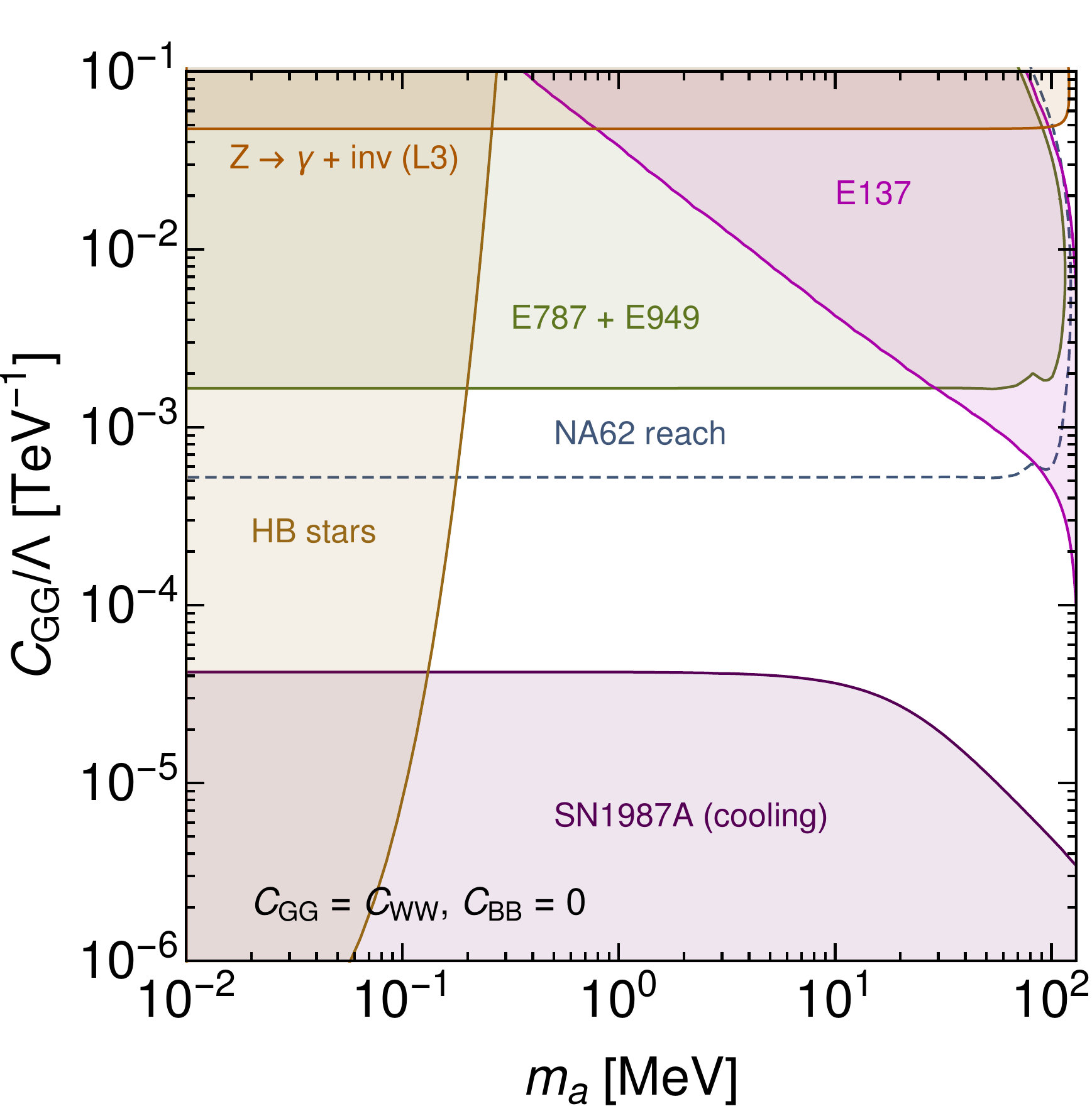}\\
\includegraphics[width=6.75cm]{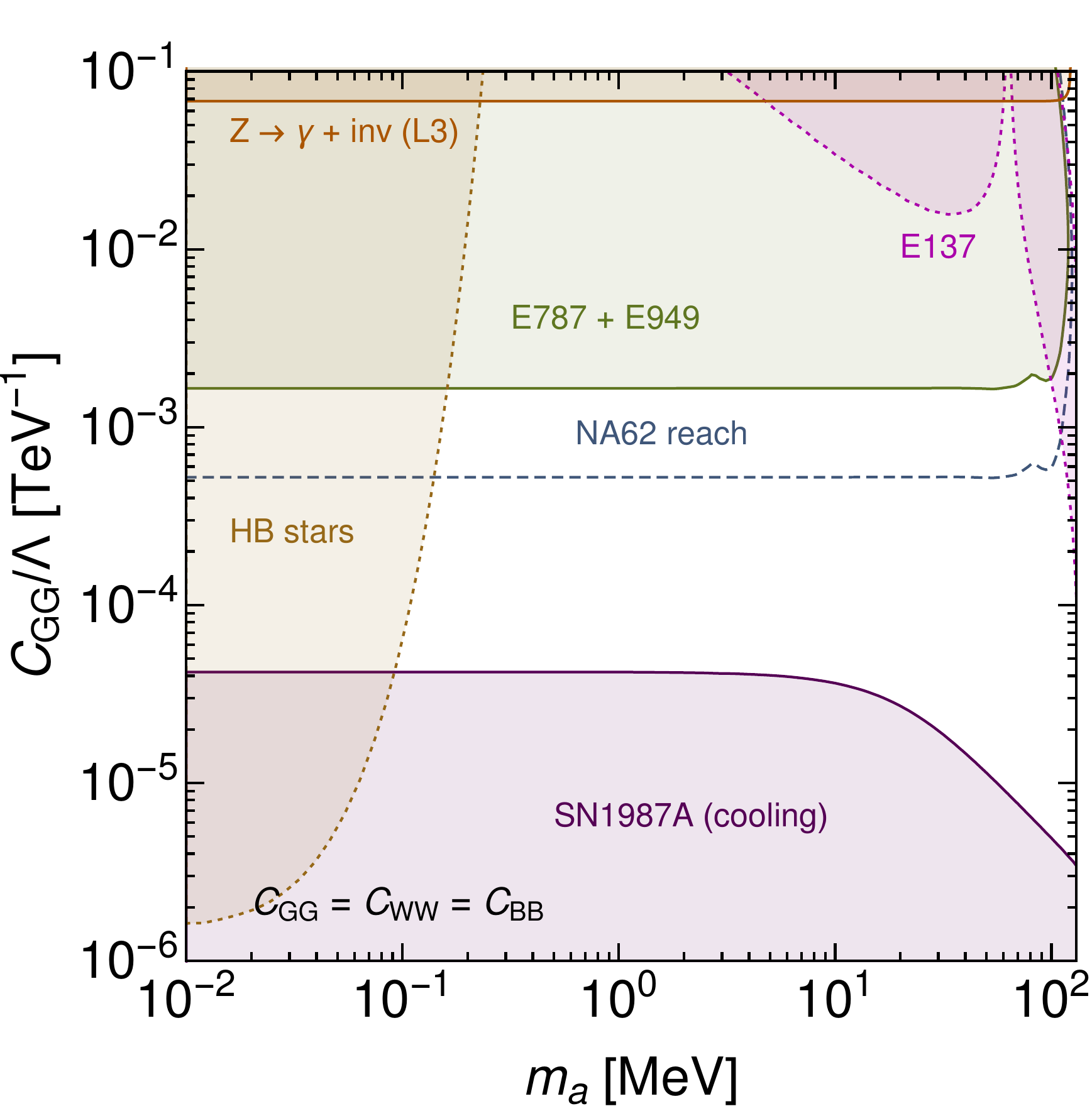}
\hspace{1cm}
\includegraphics[width=6.75cm]{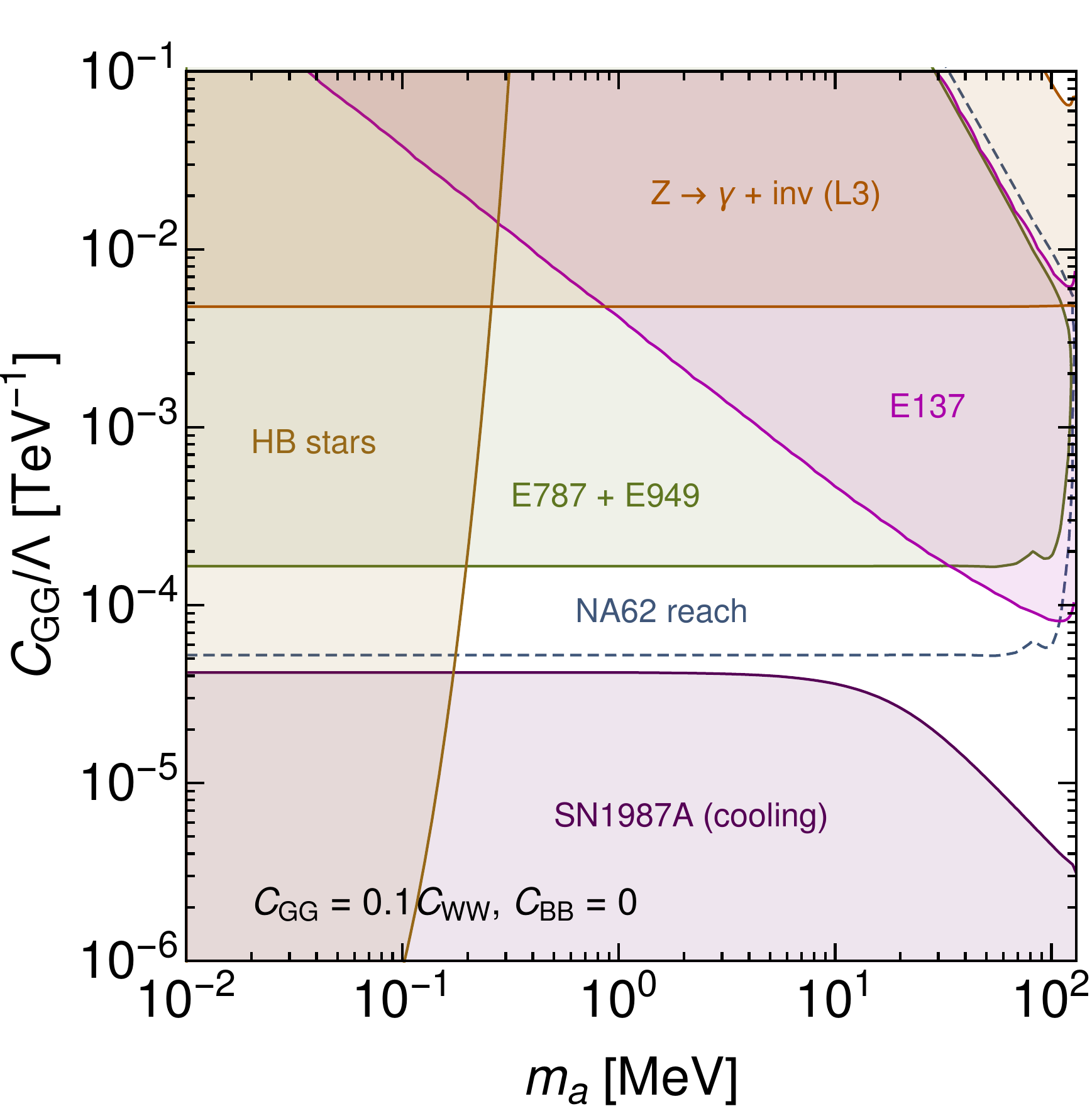}
\caption{Experimental constraints and prospects on ALPs interacting with SM gauge boson for $m_a < m_\pi$ for different ALP couplings. Starting from the top left (clockwise) we focus on dominant $C_{GG}$, then $C_{GG} = C_{WW}$ followed by $C_{GG} = 0.1\,C_{WW}$ and lastly $C_{GG} = C_{WW} = C_{BB}$. Constraints that are affected by hadronic uncertainties (most importantly the ALP mixing with $\eta$ and $\eta'$) are shown with dotted lines.}\label{fig:ConstraintsEWGlu} 
\end{figure}

The interplay of different couplings becomes even more interesting when including the ALP-gluon coupling in the discussion. Figure~\ref{fig:ConstraintsEWGlu} shows the resulting constraints as a function of $m_a$ and $C_{GG}$ for four different combinations of Wilson coefficients. The figure focuses on $m_a < m_\pi$ (corresponding to signal region 1 in NA62), while the case $m_a > m_\pi$ (signal region 2) is discussed in appendix~\ref{app:NA62SR2}. The top-left panel corresponds to the case where only the gluon coupling is relevant and all other couplings can be neglected. As discussed in section 2, the low-energy phenomenology in this case is determined by the effective ALP-photon coupling, the effective ALP-nucleon coupling and the FCNCs induced by meson mixing. The ALP-photon coupling leads to relevant constraints from E137 and HB stars. The constraint from SN1987A depends on both the photon and the nucleon coupling, but the latter dominates the phenomenology because the high nuclear density in a supernova core leads to an enhancement of bremsstrahlung processes over Primakoff conversion. As a result, the cooling constraints are sensitive to much smaller Wilson coefficients than for the case of ALPs coupling to electroweak gauge bosons. At the same time, the trapping regime extends to much smaller couplings, opening up large regions of unconstrained parameter space. 

For the case of ALP-gluon couplings, constraints on $K^+ \to \pi^+ a$ from E787/E949 are therefore highly relevant and the prospects of NA62 look very promising. However, it is important to keep in mind that these constraints, which are obtained from the ALP-meson mixing, are subject to hadronic uncertainties, see eq.~\eqref{eq:K2piaMix} and the surrounding text. It is therefore interesting to consider the case where the effective coupling to $W$ bosons gives a relevant contribution to the flavour-changing processes. One example is shown in the top-right panel, which corresponds to the case $C_{GG} = C_{WW}$ (keeping $C_{BB} = 0$). In this case the constraint from E787/E949 shifts downward and become largely independent of hadronic uncertainties. At the same time the constraint from E137 shifts upwards, because $C_{GG}$ and $C_{WW}$ contribute to the effective photon coupling with opposite sign and therefore cancel partially. The bound from SN1987A is unaffected by the ALP-photon coupling and therefore stays the same in all panels. Furthermore, we now also obtain constraints from L3, which rules out additional parameter space for $m_a \approx 100\,\mathrm{MeV}$.

In the bottom-left panel, where we take $C_{GG} = C_{WW} = C_{BB}$, we observe an approximate cancellation in the effective photon coupling (see eq.~\eqref{eq:EffPhotCoupl}) leading to a strong suppression of the bound from E137 and moving the HB star bound to larger couplings. Moreover, the E137 bound now depends sensitively on the contribution to the effective photon coupling from ALP-meson mixing, which leads to a more complicated dependence on the ALP mass and larger hadronic uncertainties. For this specific combination of Wilson coefficients the constraints on the parameter space resulting from the ALP-photon interaction hence become less important.  We point out that it is quite plausible that the Wilson coefficients $C_{GG}$, $C_{WW}$, and $C_{BB}$ would be comparable in size, for example if they are generated from the contribution of heavy new particles through triangle diagrams~\cite{Alonso-Alvarez:2018irt}. 

As an alternative, we consider in the bottom-right panel a more hierarchical coupling structure, with $C_{GG} = 0.1 C_{WW}$ and $C_{BB} = 0$. As expected, the bounds from E787/E949 and L3 become significantly stronger in this case and the sensitivity of NA62 extends almost down to the parameter region excluded by SN1987A, such that the uncertainties in the SN bound become particularly relevant. For even larger hierarchy between $C_{GG}$ and $C_{WW}$ it may even be possible to explore the entire trapping regime with laboratory experiments. The key point, however, is that the individual bounds depend in different ways on the ALP couplings and can therefore shift relative to each other. A simple one-operator study is insufficient to capture all of these dependencies and may give a misleading impression of the status of ALP models. A more careful analysis instead reveals interesting regions of parameter space where ALPs may be discovered with ongoing or future experiments.

\section{Conclusions}
\label{sec:conclusions}

The interest in the study of hidden sectors, i.e.\ new light particles with extremely weak interactions, has increased substantially over recent years as a reaction to the absence of evidence for new physics at the electroweak scale. The growing theoretical and experimental efforts have paid particular attention to ALPs, which can arise as Pseudo-Nambu-Goldstone bosons from a broken global symmetry. In the present work we have studied MeV-scale ALPs in an EFT set-up with a particular focus on couplings to SM gauge bosons. We improve on the common approach to focus on a single effective operator at a time by taking into account all relevant low-energy interactions and their correlations. Relevant effects are found to stem from ALP interactions with photons, $W\pm$ bosons and nucleons, as well as ALP-meson mixing. We study the interplay of the resulting constraints and map out the phenomenology of ALPs with several types of interactions.

We are particularly interested in the prospects of searching for MeV-scale ALPs in the laboratory by performing measurements in the $K^+ \rightarrow \pi^+ + \text{inv.}$ channel, for which the presently running NA62 experiment promises to significantly improve experimental sensitivity. We compare the projected reach of NA62 with complementary constraints on the ALP parameter space, most importantly from SN1987A. We improve upon existing calculations by consistently including both ALP-photon and ALP-nucleon couplings and estimating the effects of systematic uncertainties on both bounds. As shown in figure~\ref{fig:ComparisonPhotSN}, our main constraint is more conservative in the trapping regime for ALP-photon couplings, while for ALP-gluon couplings our results exclude more parameter space for larger ALP masses. We furthermore discuss the details of ALP-meson mixing and estimate the resulting contribution to flavour-changing processes.

As a first application of our framework we have focussed on ALPs that interact dominantly with electroweak gauge bosons. The case that ALPs couple only to $SU(2)_L$ gauge bosons is already severely constrained and almost the entire parameter space relevant for NA62 is already excluded. However, as soon as a simultaneous coupling to hypercharge gauge bosons is considered, the different constraints can be shifted relative to each other. In particular, by enhancing the coupling to photons it is possible to extend the trapping regime and circumvent the cooling bound from SN1987A, see figure~\ref{fig:ConstraintsEW}. 

We then studied the entire range of interactions with gauge bosons by also considering the ALP-gluon coupling. The resulting ALP-nucleon interactions push the cooling constraint from SN1987A to much smaller couplings, opening up large regions of unconstrained parameter space that can be explored with laboratory experiments like NA62 (figure~\ref{fig:ConstraintsEWGlu}). By varying the relative magnitude of the different Wilson coefficients, we illustrate how the individual bounds and prospects (and hence the overall phenomenology) depend on the different couplings.

Our results demonstrate that it is essential to simultaneously consider multiple effective ALP interactions, which are generically expected to be generated simultaneously in many ultraviolet completions. Their correlations have an important influence on the interplay of different probes and the viable regions of parameter space. Focusing on single operators may introduce significant biases, for example by overestimating the strength of certain exclusion limits. Nevertheless, we still find that there are many strong constraints on the ALP parameter space, which can be improved with further theoretical and experimental efforts. In particular, it would be very interesting to include MeV-scale ALPs in a supernova simulation to study their impact in more detail and obtain more robust constraints from SN1987A. Likewise, we have left a detailed study of constraints and sensitivity projections for proton beam dump experiments to future work.

\acknowledgments

We thank Gonzalo Alonso-\'{A}lvarez, Babette D\"{o}brich, Matthew Dolan, Frederick Hiskens, Joerg Jaeckel, Georg Raffelt, Tommaso Spadaro, Lennert Thormaehlen and Susanne Westhoff for fruitful discussions and Samuel McDermott for helping us with the implementation of supernova constraints. 
This  work  is  funded  by  the  Deutsche Forschungsgemeinschaft (DFG) through the Emmy Noether Grant No.\ KA 4662/1-1 and the Collaborative Research Center TRR 257 ``Particle Physics Phenomenology after the Higgs Discovery''.


\appendix
\section{Details on ALP interactions with hadrons}
\label{app:ChiralInt}
In this appendix we provide details on the computation of the ALP mixing with pseudoscalar mesons. We start from the second line of eq.~\eqref{eq:LagrChirRot} and map it onto a chiral Lagrangian at leading order. This yields~\cite{Kaiser:1998ds,Kaiser:2000gs,Bickert:2016fgy,Scherer:2012xha}
\begin{align}
 \begin{split}
\label{eq:LOChPTLagr}
\mathcal{L}_{\chi PT}  = \frac{F_0^2}{4} \text{Tr}[(D_\mu U) (D^\mu U)^\dagger]+ \frac{F_0^2 B_0}{2} \text{Tr}[M_q(a) U^\dagger + U M^\dagger_q(a)] -\frac{1}{2} M_0 \eta_0^2\,.
\end{split}
 \end{align}
Here $M_0$ accounts for the mass term induced by the $U(1)_A$ breaking, $B_0 = m_{\pi}^2/(m_u + m_d)$ and $U = e^{i \phi/F_0}$ with the matrix $\phi$ containing the fields of the pseudoscalar mesons
\begin{align}
\phi =
\begin{pmatrix}
  \pi^0 + \frac{1}{\sqrt{3}} \eta_8 & \sqrt{2} \pi^+ & \sqrt{2} K^+ \\
  \sqrt{2} \pi^- & -\pi^0 + \frac{1}{\sqrt{3}} \eta_8 & \sqrt{2} K^0 \\
  \sqrt{2} K^- & \sqrt{2} \bar{K}^0 &  -\frac{2}{\sqrt{3}} \eta_8 \\
\end{pmatrix}
+ \sqrt{\frac{2}{3}} \eta_0\,\text{diag}(1,1,1)\,.
\end{align}
The mass matrix is given by $M_q(a) = e^{-i \kappa_q a/2 f_a} M_q e^{-i \kappa_q a/2 f_a}$ with \mbox{$1/f_a = -32\pi^2 C_{GG}/\Lambda$}. We choose $\kappa_q = M^{-1}_q / \text{Tr}[M^{-1}_q]$ to ensure that there is no mass mixing between the ALP and the octet mesons~\cite{Georgi:1986df}. The covariant derivative is defined by
\begin{align}
 \begin{split}
  D_\mu U = \partial_\mu U - i (v_\mu + a_\mu) U + i U (v_\mu - a_\mu)\,,
  \end{split}
 \end{align}
with the spurions reading
\begin{align}
 \begin{split}
  v_\mu &= - e Q_q A_\mu\,,\\
  a_\mu &= \frac{\partial_\mu a}{2 \Lambda} 32 \pi^2 C_{GG} \kappa_q\,.
  \end{split}
 \end{align}
We first rotate $\eta_8$ and $\eta_0$ into $\eta$ and $\eta'$
\begin{align}
 \begin{split}
  \begin{pmatrix}
   \eta_8\\
   \eta_0
  \end{pmatrix}
=
\begin{pmatrix}
 c(\theta) & s(\theta)\\
 -s(\theta) & c(\theta)
\end{pmatrix}
  \begin{pmatrix}
   \eta\\
   \eta'
  \end{pmatrix}
  \,,
  \end{split}
 \end{align}
where we have introduced the short-hand notation $c(\theta) \equiv \cos(\theta) $ and $ s(\theta) \equiv \sin(\theta)$. We adopt the value $\theta = -13^{\circ}$ from ref.~\cite{Bossi:2008aa}. To diagonalize the $\eta\text{--}\eta'$ sub-mass-matrix, we then have to take $M_0 \approx 1.05\,\mathrm{GeV}$ resulting in $m_\eta \approx 537\,\mathrm{MeV}$ and $m_{\eta'} \approx 1.15\,\mathrm{GeV}$ sufficient for our intended accuracy. Expanding now eq.~\eqref{eq:LOChPTLagr}, one obtains both kinetic and mass mixing contributions. Following the notation of ref.~\cite{Aloni:2018vki}, we write these as
\begin{align}
 \begin{split}
  \mathcal{L} = \frac{1}{2} (\partial_\mu P_i) \,K_{ij}\, (\partial^\mu P_j) - \frac{1}{2} P_i \,M^2_{ij}\, P_j
  \end{split}
 \end{align}
with $P = (a,\,\pi^0,\,\eta,\,,\eta')$. The matrices read
\begin{align}
 \begin{split}
 K = 
    \begin{pmatrix}
   1 & - \epsilon K_{a\pi}&  - \epsilon K_{a\eta} &  - \epsilon K_{a\eta'}\\
    - \epsilon K_{a\pi}  & 1 & 0 & 0\\
    - \epsilon K_{a\eta} & 0 & 1 & 0\\
    - \epsilon K_{a\eta'} & 0 & 0 &1 \\
  \end{pmatrix}
  \,,\quad M^2 = 
      \begin{pmatrix}
   m_a^2 & 0& \epsilon\, m_{a\eta}^2 & \epsilon \,m_{a\eta'}^2\\
    0  & m_\pi^2 & \delta_I m_{\pi\eta}^2 & \delta_I\,m_{\pi\eta'}^2\\
   \epsilon\, m_{a\eta}^2 & \delta_I\, m_{\pi\eta}^2  & m_\eta^2 & 0\\
    \epsilon \,m_{a\eta'}^2 & \delta_I\, m_{\pi\eta'}^2 & 0 &m_{\eta'}^2 \\
  \end{pmatrix}
  \,,
  \end{split}
 \end{align}
where $\delta_I = (m_d - m_u)/(m_u + m_d)$ and
\begin{align}
 \begin{split}
  m^2_{\pi\eta} = - m_\pi^2 \left(\frac{c(\theta)}{\sqrt{3}} - \sqrt{\frac{2}{3}} s(\theta)\right)\,,\\
  m^2_{\pi\eta'} = - m_\pi^2 \left(\sqrt{\frac{2}{3}} c(\theta) + \frac{s(\theta)}{\sqrt{3}}\right)\,.
  \end{split}
 \end{align}
For the remaining expressions we refer to eqs.~\eqref{eq:KinMixDef} and \eqref{eq:MassMixDef}. Diagonalisation of the kinetic as well as mass matrix results in the mixing angles given in eq.~\eqref{eq:LOMix} at leading order.

\section{Mixing contributions to \texorpdfstring{$K^+ \rightarrow \pi^+ a$}{KtoPiA}}
\label{app:KtoPiMix}
The hierarchy encountered in kaon decays, namely $\Gamma(K^0 \rightarrow \pi^+ \pi^-)$, $\Gamma(K^0 \rightarrow \pi^0 \pi^0) \gg \Gamma(K^+ \rightarrow \pi^+ \pi^0)$, is usually explained by considering the leading order chiral Lagrangian for $\Delta S = 1$ transitions~\cite{Cirigliano:2003gt,Cirigliano:2011ny,Kambor:1989tz}
\begin{align}
 \begin{split}
 \label{eq:KtoPiLagr}
  \mathcal{L}_{\Delta S = 1} \supset G_8 F_0^4 \text{Tr}\left[\frac{\lambda_6 - i \lambda_7}{2} \partial^\mu U^\dagger \partial_\mu U\right] + G_{27} F_0^4 \left(L_{\mu23} L^\mu_{11} + \frac{2}{3} L_{\mu21} L^\mu_{13}\right) + \text{h.c.}
  \end{split}
 \end{align}
Here $\lambda_i$ denote the Gell-Mann matrices, $L_\mu = i U^{\dagger} D_\mu U$ and the indices attached to $L_{\mu}$ denote specific matrix elements. From this Lagrangian, we obtain for the amplitude
\begin{align}
 \begin{split}
  i \mathcal{M}(K^0 \rightarrow \pi^+ \pi^-) &= A_0 e^{i \chi_0} + \frac{1}{\sqrt{2}} A_2 e^{i \chi_2} \approx A_0 e^{i \chi_0} \approx \sqrt{2} G_8 F_0 m_K^2 \,,\\
  i \mathcal{M}(K^+ \rightarrow \pi^+ \pi^0) &=\frac{3}{2} A_2 e^{i \chi_2} \approx \frac{5}{3} G_{27} F_0 m_K^2 \,,\\
  \end{split}
 \end{align}
where $A_i$ and $\chi_i$ describe the amplitudes and phases in the isospin decomposition of the kaon amplitudes~\cite{Cirigliano:2003gt,Cirigliano:2011ny}. We have expanded each amplitude in $A_0/A_2 \approx 22$ and additionally in $m_K^2/m_\pi^2$ at leading order. As pointed out in ref.~\cite{Bardeen:1986yb} the enhanced term proportional to $G_8$ in eq.~\eqref{eq:KtoPiLagr} also contributes to the off-shell amplitudes
\begin{align}
 \begin{split}
i \mathcal{M}(K^0 \rightarrow \pi^+ \eta) & = \frac{G_8 F_0}{\sqrt{3}} \left(2m_K^2 \big(c(\theta) - \sqrt{2} s(\theta)\big) + m_\pi^2 \big(c(\theta) +2 \sqrt{2} s(\theta)\big) -3 p_\eta^2 c(\theta) \right)\\
  &\approx \frac{G_8 F_0}{\sqrt{3}} 2 m_K^2 \big(c(\theta) - \sqrt{2} s(\theta) \big)\,,\\
   i \mathcal{M}(K^0 \rightarrow \pi^+ \eta') &= \frac{G_8 F_0}{\sqrt{3}} \left(2m_K^2 \big(  \sqrt{2} c(\theta) +s(\theta)\big) - m_\pi^2 \big(2 \sqrt{2}c(\theta) - s(\theta)\big) -3 p_{\eta'}^2 s(\theta) \right)\\
  &\approx \frac{2G_8 F_0}{\sqrt{3}} m_K^2 \big(  \sqrt{2} c(\theta) +s(\theta)\big)\,.
  \end{split}
 \end{align}
These amplitudes are enhanced compared to $i \mathcal{M}(K^+ \rightarrow \pi^+ \pi^0)$ and are therefore also included. The $K^+ \rightarrow \pi^+ a$ amplitude can then be determined by taking all mixing contributions into account
\begin{align}
\notag
 i & \mathcal{M}(K^+ \rightarrow \pi^+ a) \\\notag
 & \approx \theta_{a\pi}\, i \mathcal{M}(K^+ \rightarrow \pi^+ \pi^0) + \theta_{a\eta}\, i \mathcal{M}(K^+ \rightarrow \pi^+ \eta) +  \theta_{a\eta'}\, i \mathcal{M}(K^+ \rightarrow \pi^+ \eta')\\
 \notag
 &\approx \Big[\theta_{a\pi} \frac{3A_2}{2A_0} e^{i(\chi_2 - \chi_0)}  + \theta_{a\eta} \sqrt{\frac{2}{3}} \big(c(\theta) - \sqrt{2} s(\theta)\big) + \theta_{a\eta'} \sqrt{\frac{2}{3}} \big(\sqrt{2} c(\theta) + s(\theta)\big)\Big] i\mathcal{M}(K^0 \rightarrow \pi^+ \pi^-)\\
 & =\theta_\text{mix} \,i \mathcal{M}(K^0 \rightarrow \pi^+ \pi^-)\,,
 \end{align}
coinciding with eq.~\eqref{eq:KtoPiMixAmpl}.

\section{Details on the computation of the ALP luminosity in SN1987A}
\label{app:LumDetails}
The general expression for the ALP luminosity involves a volume integral over the energy loss rate $Q$, where also the probability $e^{-\tau}$ of an ALP escaping the neutrinosphere is taken into account~\cite{Chang:2016ntp,Chang:2018rso}:
\begin{align}
 L = \int_{r \leq R_\nu} dV\, Q\, e^{-\tau}\,.
 \end{align}
Here the energy loss rate $Q$ (energy per volume and unit time) is defined by~\cite{Raffelt:1990yz}
\begin{align}
 \begin{split}
\label{eq:Qrate}
Q = &\int \frac{\text{d}^3\textbf{p}_a}{2 \omega_a (2\pi)^3} \omega_a \left(\prod_i \int \frac{\text{d}^3\textbf{p}_i}{2\omega_i (2\pi)^3} f_i(\omega_i)\right) \left(\prod_f \int  \frac{\text{d}^3\textbf{p}_f}{2\omega_f (2\pi)^3} [ 1\pm f_f(\omega_f)]\right)\\
&\times \,S \sum_{\text{spins/pol.}} |M|^2\,(2\pi)^4 \,\delta^{(4)}\Big(\sum_i p^\mu_i - \sum_f p^\mu_f - p^\mu_a\Big)\,,
\end{split}
\end{align}
where $\textbf{p}_j$ ($\omega_j$) denote the three-momenta (energies) of the ALP ($j = a$), the initial states ($j = i$) and final states ($j = f$). Moreover, $f(\omega_j)$ are the phase-space occupation numbers defined by $n_j = g_j\int \text{d}^3 \textbf{p} f(\omega_j)/(2\pi)^3 $ with the number density $n_j$ and degeneracy factor $g_j$ of the particle species $j$. Note that a factor of $1 + f(\omega_j)$ is taken for bosonic final states (stimulated emission), whereas $1-f(\omega_j)$ applies to fermions (Pauli blocking). Furthermore, $S$ denotes a symmetry factor for identical states in the initial or final state and $|M|^2$ is the squared matrix element which is summed over initial and final state spins and polarizations.

The optical depth $\tau$ is defined by an integral over the inverse mean free path $l = \beta/ \Gamma_\text{abs}$~\cite{Raffelt:1988rx,Raffelt:1990yz}:
\begin{align}
  \tau = \int \text{d}r \beta^{-1} \Gamma_\text{abs} \,,
 \end{align}
with the absorption rate $ \Gamma_\text{abs}$ for a process $a + i \rightarrow f$ reading~\cite{Weldon:1983jn}
\begin{align}
 \begin{split}
  \Gamma_\text{abs} = &\frac{1}{2\omega_a} \left(\prod_i \int \frac{\text{d}^3\textbf{p}_i}{2\omega_i (2\pi)^3} f_i(\omega_i)\right) \left(\prod_f \int \frac{\text{d}^3\textbf{p}_f}{2\omega_f (2\pi)^3} [ 1\pm f_f(\omega_f)]\right)\\
  &\times S \sum_{\text{spins/pol.}} |M|^2\,(2\pi)^4 \,\delta^{(4)}\Big(\sum_i p^\mu_i - \sum_f p^\mu_f - p^\mu_a\Big)\,.
  \end{split}
 \end{align}
We stress that the ALP energy integration in eq.~\eqref{eq:Qrate} also affects the optical depth. In the case of the Primakoff process, where one integrates over the photon momentum in eq.~\eqref{eq:LumFormula}, we exploit that the photon energy is equal to the ALP energy.

The different rates have already been computed in the literature and simply need to be combined. For ALP production through bremsstrahlung in nucleon scattering and Primakoff processes we have~\cite{Raffelt:2006cw,Raffelt:1996wa}
\begin{align}
 \begin{split}
 \label{eq:EnergyRates}
  Q_\text{brems} = \int \frac{\text{d}^{3}\textbf{p}_a}{(2\pi)^3} \omega  \Gamma_a e^{-\omega/T} \beta\,,\\
  Q_\text{prim} = \int \frac{2\text{d}^{3} \textbf{p}_\gamma}{(2\pi)^3} \frac{\omega \Gamma_{\gamma \rightarrow a}}{e^{\omega/T}-1} \beta\,,
  \end{split}
 \end{align}
where $\omega_a \equiv \omega$ denotes the ALP/photon energy and the phase space factors $\beta$ account for sizeable ALP masses. With these results, we can also determine the different absorption rate formulas\footnote{In refs.~\cite{Dolan:2017osp,Cadamuro:2011fd} an additional $\beta$ factor occurs in the definition of the mean free path for the Primakoff process as $\Gamma_{\gamma\rightarrow a}$ for massive ALPs is used. In this case one should also replace $\Gamma_{\gamma\rightarrow a}\,\beta$ in the emission rate in eq.~\eqref{eq:EnergyRates} by the formula for $\Gamma_{\gamma\rightarrow a}$ for massive ALPs to be consistent. We find that both approaches lead to similar results.}
\begin{align}
 \begin{split}
  \Gamma_\text{abs}^\text{brems} &= \Gamma_a\,,\\
  \Gamma_\text{abs}^\text{prim} &= 2 \Gamma_{\gamma\rightarrow a}\,,\\
  \Gamma_\text{abs}^\text{decay} &= \gamma^{-1} \Gamma_{a\gamma\gamma}\,,
  \end{split}
 \end{align}
where an inverse Lorentz factor $\gamma = \omega / m_a$ is included in the last line as $\Gamma_{a\gamma\gamma}$ is computed in the ALP rest frame. Note that we neglected the Bose factors for the photon final states for simplicity as these barely influence the results.

\begin{figure}[t]
\centering
\includegraphics[width=6.75cm]{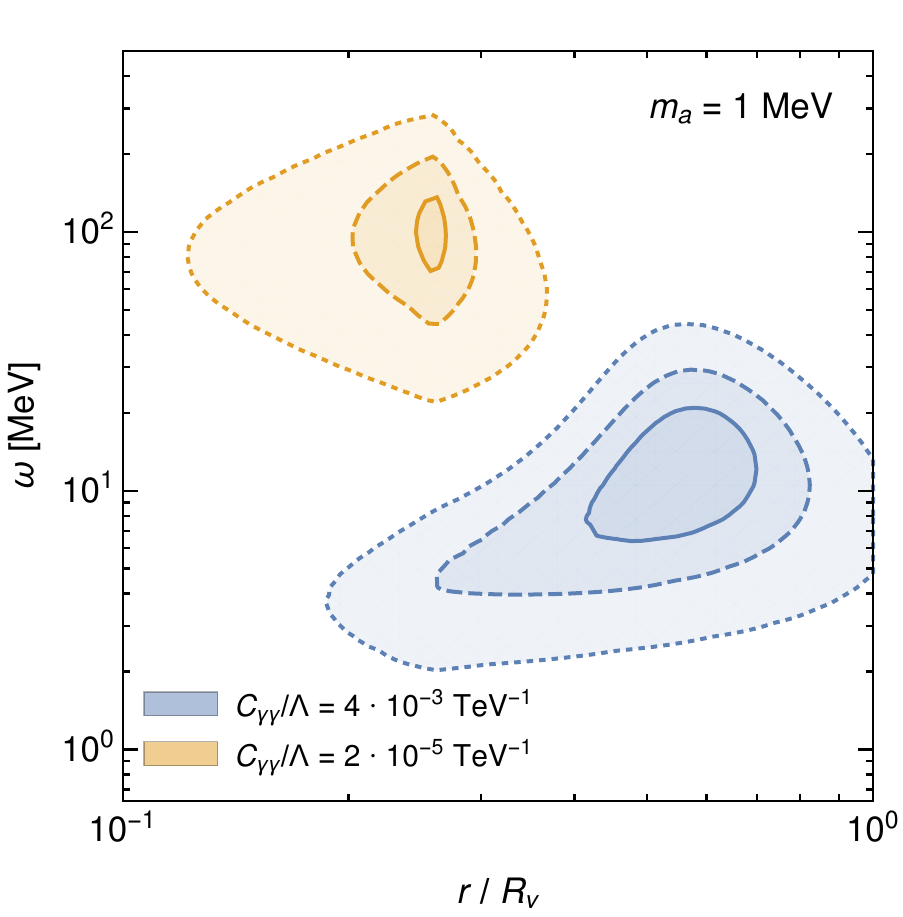}
\hspace{1cm}
\includegraphics[width=6.75cm]{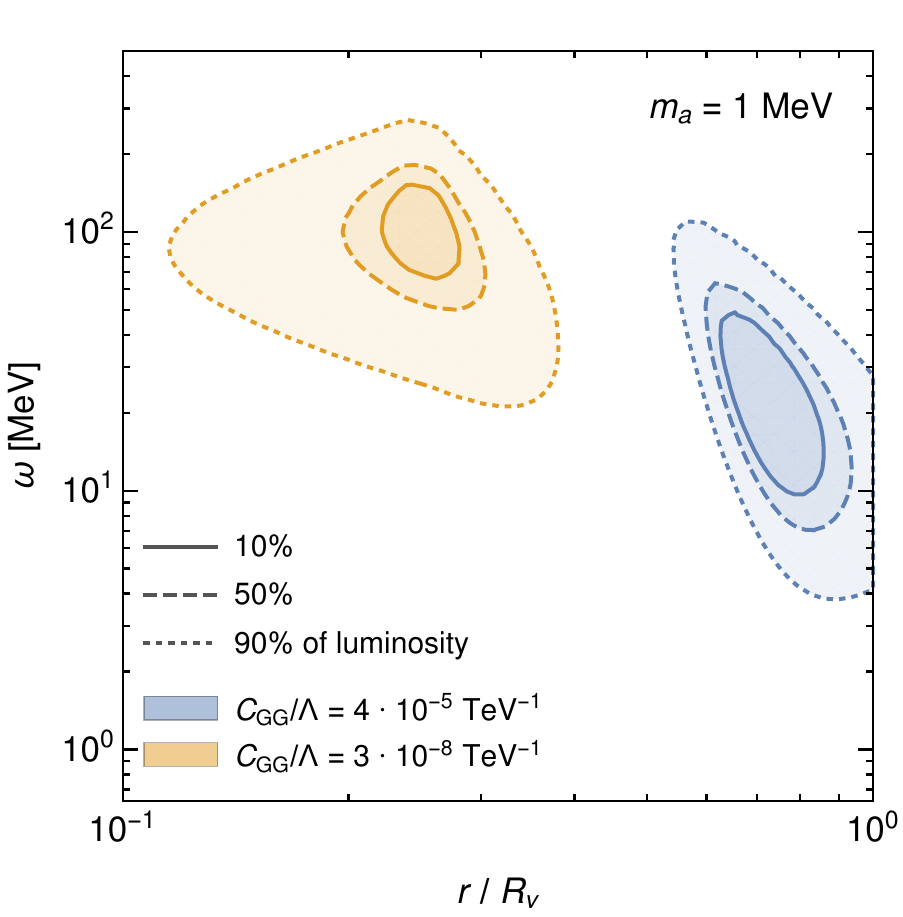}
\caption{Luminosity distribution of ALP energies as a function of radius for $m_a = 1\,\mathrm{MeV}$. On the left we focus on a dominant ALP-photon coupling and on the right on an ALP-gluon coupling.}\label{fig:LumDistr}
\end{figure}

To illustrate the process of ALPs cooling the proto-neutron star, we provide in figure~\ref{fig:LumDistr} the distribution of ALP energies contributing dominantly to the luminosity as a function of radius for ALP-photon couplings (left) and ALP-gluon couplings (right). Here we take $m_a = 1\,\mathrm{MeV}$ and consider two different coupling values in each case, corresponding to the upper and lower edge of the excluded parameter region. We make the following observations: For smaller coupling values most of the ALPs escaping the neutrinosphere get produced in the core region, where both temperature and density are largest, and therefore also have energies of up to $\mathcal{O}(100\,\mathrm{MeV})$. On the other hand, those ALPs with rather large couplings are only able to leave the neutrinosphere when they are produced close to its edge as the trapping in the inner part is too strong. But even these ALPs have energies $\gtrsim 10\,\mathrm{MeV}$ on average for the fiducial profile, making it clear why the supernova cooling bounds in figure~\ref{fig:ComparisonPhotSN} become independent of the ALP mass for $m_a \lesssim 10\,\mathrm{MeV}$. We also note that for the other profiles discussed in appendix~\ref{app:SNUncer} the luminosity distribution in the trapping regime is confined to a smaller radial region, i.e.~resembles the emission of ALPs from a sphere, as the densities drop off more sharply.

\section{Uncertainties affecting the SN1987A cooling bounds}
\label{app:SNUncer}
\begin{figure}[t]
\centering
\hspace{-0.4cm}
\includegraphics[width=6.9cm]{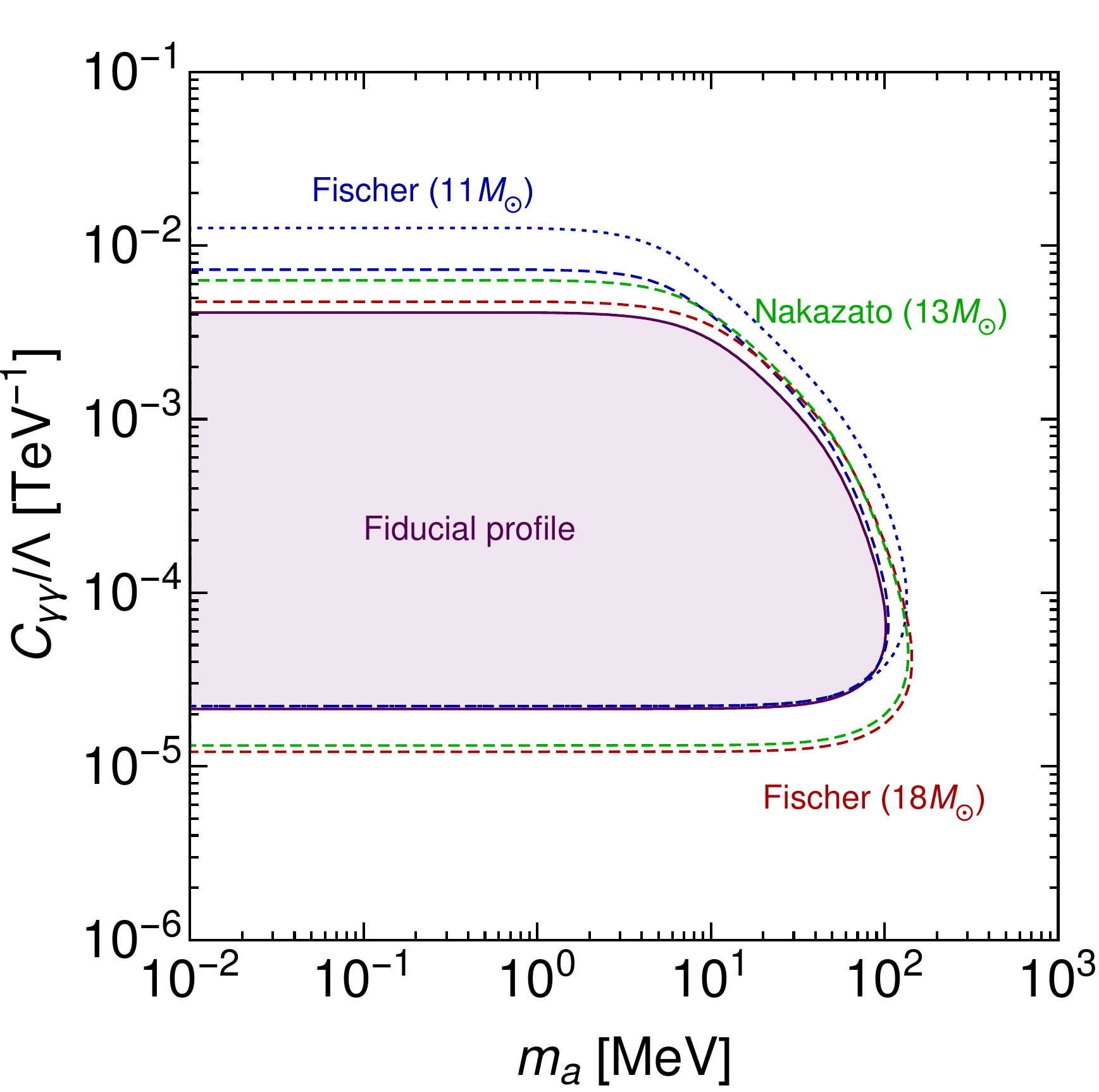}
\hspace{1cm}
\includegraphics[width=6.9cm]{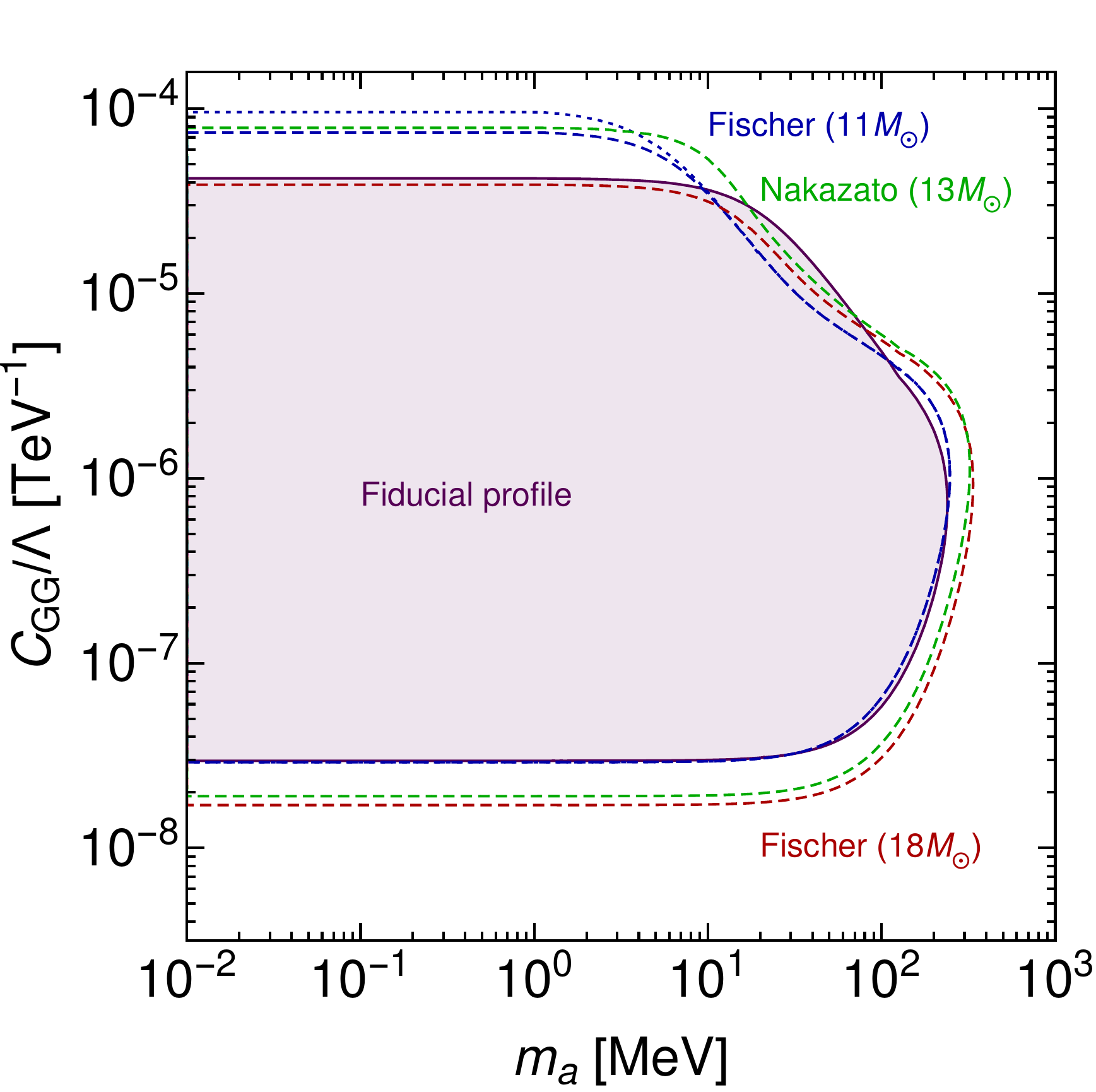}
\hspace{-0.1cm}
\caption{Comparison of the different supernova bounds for a dominant ALP-photon coupling (left) and a dominant ALP-gluon interaction (right). In both panels we varied the temperature and density profiles within our approach. We used those from Fischer ($11\,M_{\odot}$ and $18\,M_{\odot}$)~\cite{Fischer:2016cyd} as well as Nakazato ($13\,M_{\odot}$)~\cite{Nakazato:2012qf} and the purple coloured area corresponds to the fiducial profile used in the main text. The blue dotted line correponds to Fischer ($11\,M_{\odot}$) with a significant reduction of $R_\text{far}$ down to $R_\nu + 1\,\mathrm{km} = 25.9\,\mathrm{km}$.}\label{fig:ComparisonPhotSNOwn}
\end{figure}

In this appendix we provide more details on the uncertainties affecting the SN bounds. One uncertainty results from the unknown temperature and density profiles of the underlying star which can lead to sizeable differences, see figure~\ref{fig:ComparisonPhotSNOwn}. We use the profiles from refs.~\cite{Fischer:2016cyd,Nakazato:2012qf} with the parameters as given in table 2 of ref.~\cite{Chang:2016ntp} and set $R_\text{far} = 80\,\mathrm{km}$ such that ALPs have to traverse roughly the same radial distance outside of $R_\nu$ as for the fiducial profile. Another source for differences is the precise value of $R_\text{far}$. As an example, we also include the bounds based on the Fischer ($11\,M_{\odot}$) profile by reducing $R_\text{far}$ significantly to \mbox{$R_\text{far} = R_\nu + 1\,\mathrm{km} = 25.9\,\mathrm{km}$}. Similar reaches in the trapping regime can also be obtained for the other three profiles when reducing $R_\text{far}$. On the other hand, increasing the neutrino luminosity to $L_\nu = 5\cdot 10^{52}\,\text{erg}/\text{s}$, as e.g.~done in ref.~\cite{Lee:2018lcj}, would relax the constraints again. Varying all of these effects simultaneously leads to sizeable variations in the bounds.

We conclude that SN cooling bounds are affected by substantial uncertainties. Since the bounds based on the fiducial profile shown in figure~\ref{fig:ComparisonPhotSN} are the most conservative ones, we choose these bounds as the main constraints in the present work. We note that there are further uncertainties that would affect the cooling bounds, such as the axion feedback to the SN explosion. To study the impact of these effects would however require dedicated SN simulations, and we therefore do not include them in the present work.

\section{Prospects of NA62 for signal region 2}
\label{app:NA62SR2}

\begin{figure}[t]
\centering
\includegraphics[width=6.75cm]{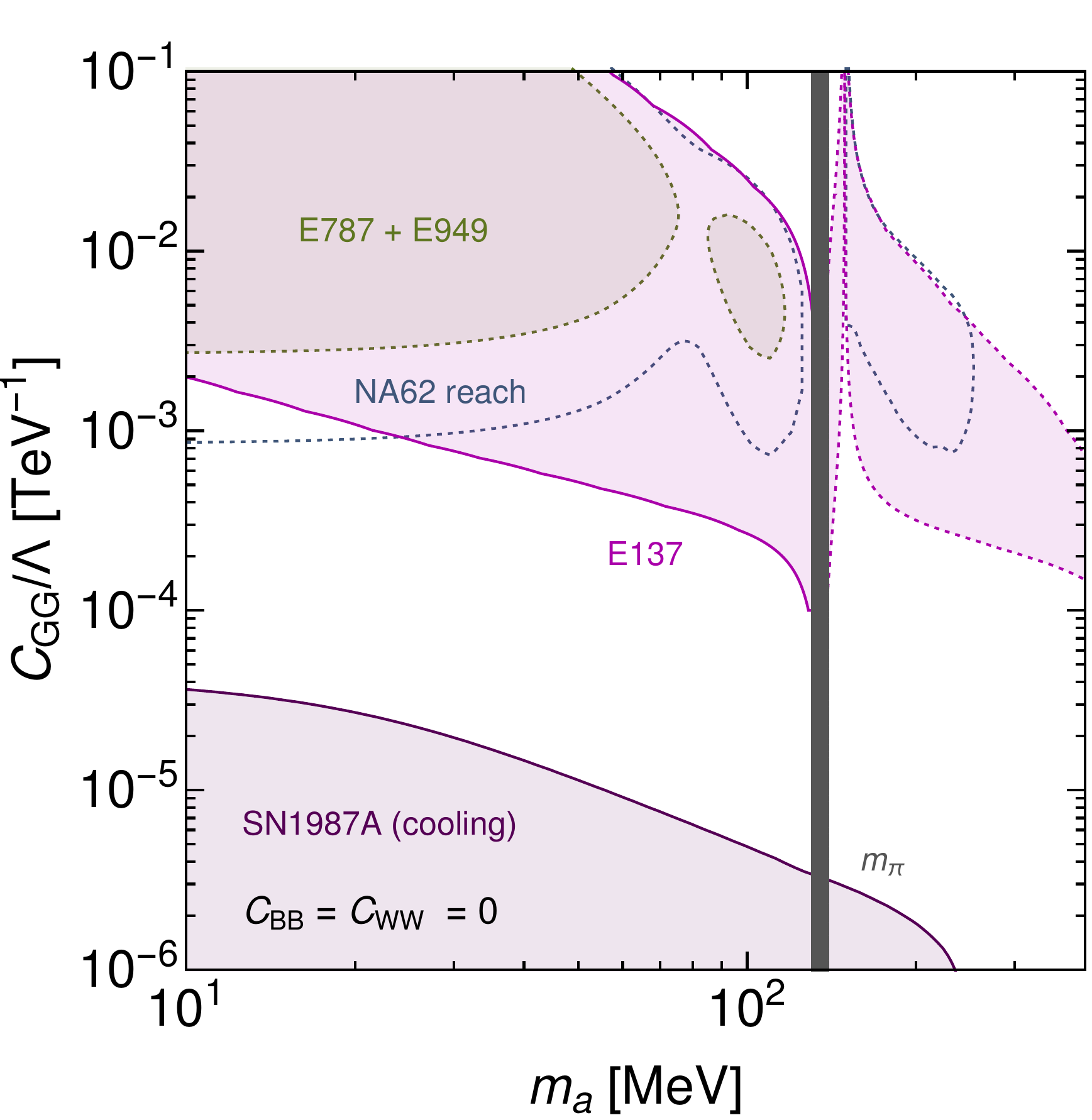}
\hspace{1cm}
\includegraphics[width=6.75cm]{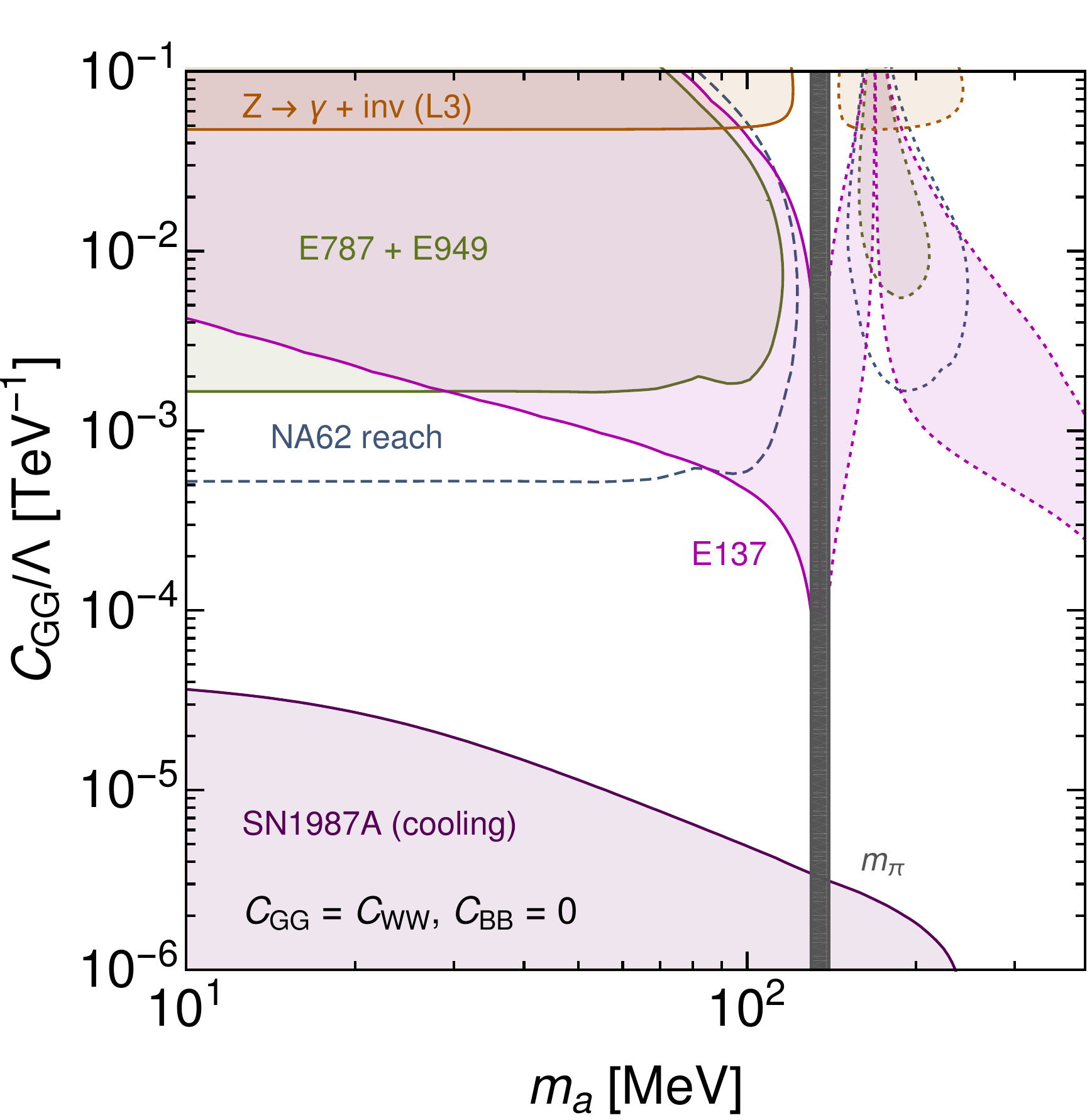}\\
\includegraphics[width=6.75cm]{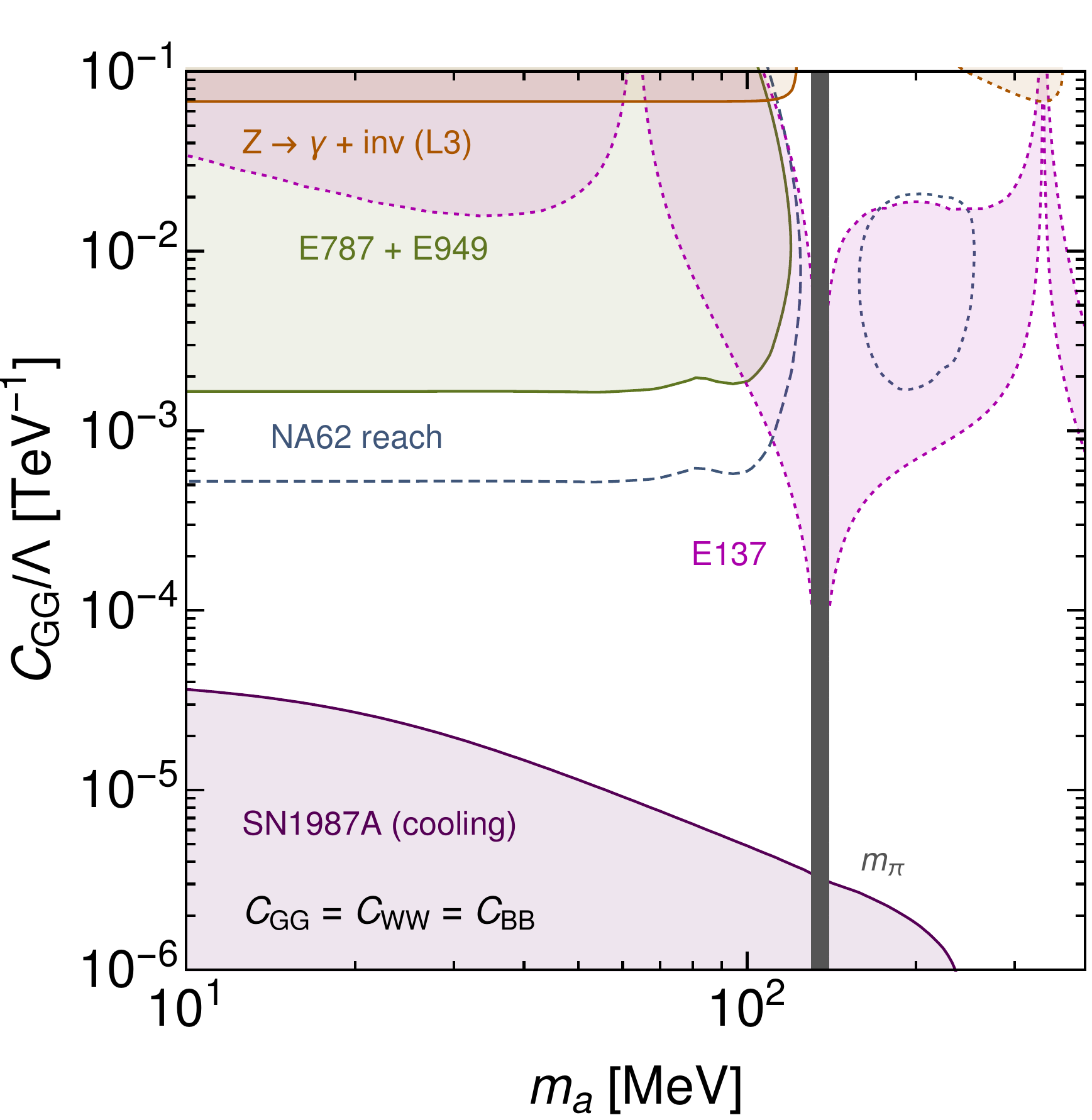}
\hspace{1cm}
\includegraphics[width=6.75cm]{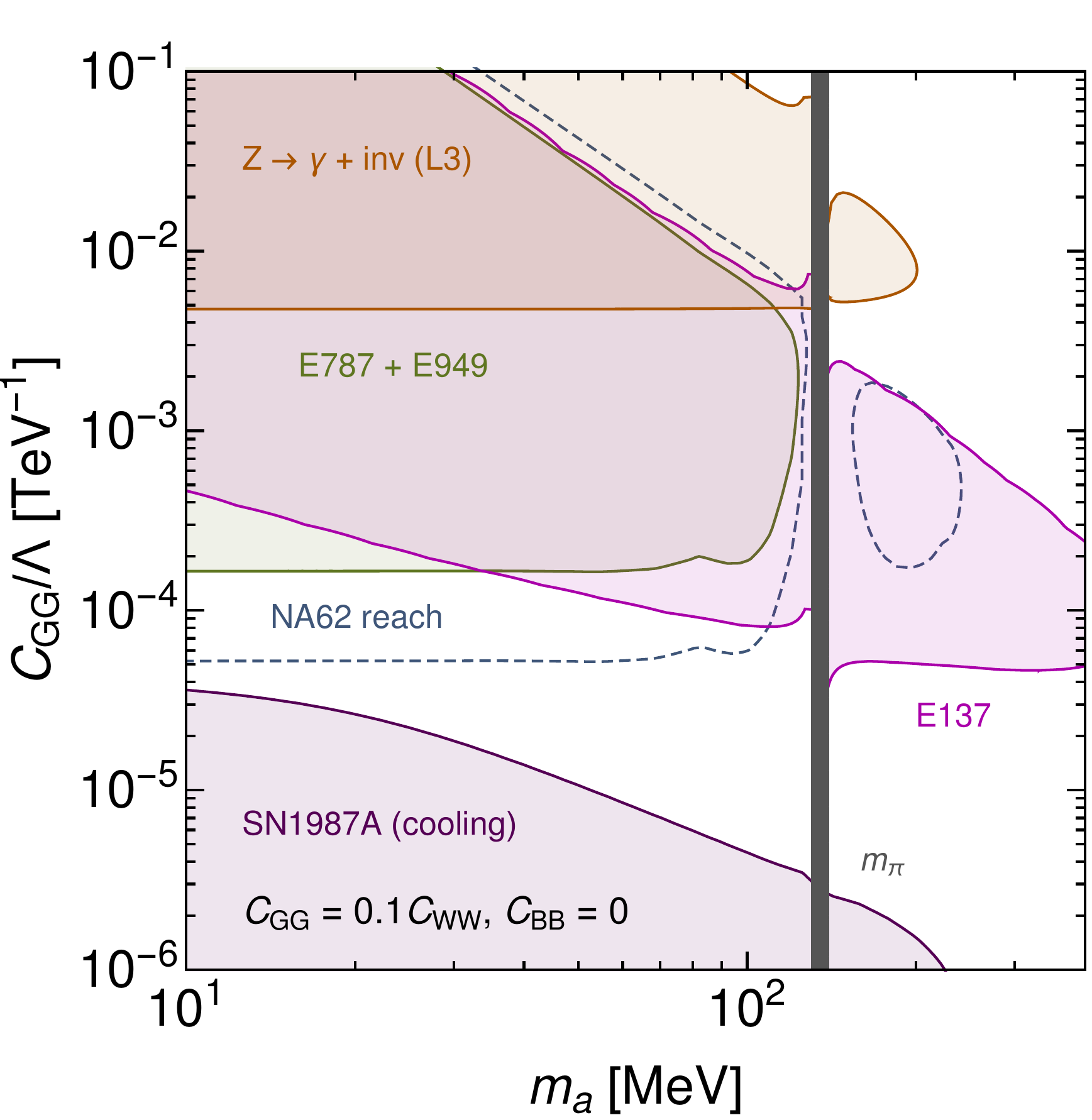}
\caption{Same as figure~\ref{fig:ConstraintsEWGlu} but including the signal region 2 of NA62. Starting from the top left (clockwise) we focus on dominant $C_{GG}$, then $C_{GG} = C_{WW}$ followed by $C_{GG} = 0.1\,C_{WW}$ and lastly $C_{GG} = C_{WW} = C_{BB}$. Constraints that are affected by hadronic uncertainties (most importantly the ALP mixing with $\eta$ and $\eta'$) are shown with dotted lines. This affects in particular signal region 2 through the effective photon coupling, except in the bottom-right panel, where the $a\text{--}\eta$ and $a\text{--}\eta'$ mixing contributions to the photon coupling are small.}\label{fig:ResultEWGApp}
\end{figure}

In this appendix we provide the constraints and prospects for the second signal region in NA62, which is sensitive to $m_a > m_\pi$. As one can see from figure~\ref{fig:ResultEWGApp}, this signal region is not as promising as the first one since it is largely excluded by E137 and L3 combined. Moreover, the general structure of the bounds here is more complex, because the mixing contributions between the ALP and pseudoscalar mesons become crucial and lead to cancellations in the ALP-photon coupling. We also point out that non-resonant searches for ALPs at the LHC can play a potential role in this mass region~\cite{Gavela:2019cmq}.

\providecommand{\href}[2]{#2}\begingroup\raggedright\endgroup

\end{document}